\documentclass[%
reprint,
superscriptaddress,
 amsmath,
 amssymb,
 aps,
prx,
floatfix,
]{revtex4-2}

\usepackage{float}
\usepackage{braket}
\usepackage{graphicx}% Include figure files
\usepackage{dcolumn}% Align table columns on decimal point
\usepackage{bm}% bold math
\usepackage{inputenc}
\usepackage{blindtext}
\usepackage{color}
\usepackage{booktabs,siunitx}
\usepackage{comment}
\usepackage{todonotes}
\usepackage{hyperref}% add hypertext capabilities
\hypersetup{colorlinks,citecolor=blue}

\graphicspath{{Figures/}}
\newcolumntype{~}{>{\global\let\currentrowstyle\relax}}
\newcolumntype{^}{>{\currentrowstyle}}

\DeclareMathAlphabet\mathbfcal{OMS}{cmsy}{b}{n}

\begin{document}

%% commandes perso
\newcommand{\JPcomment} [1] 
{\todo[inline,backgroundcolor=green,size=\small ,bordercolor=white]{{\bf JP:} #1}}
\newcommand{\ACcomment} [1] 
{\todo[inline,backgroundcolor=cyan,size=\small ,bordercolor=white]{{\bf Alo\"{i}s:} #1}}

\newcommand{\ollecomment} [1] 
{\todo[inline,backgroundcolor=red,size=\small ,bordercolor=white]{{\bf Olle:} #1}}

\newcommand{\liou}{\mathcal{L}}
\newcommand{\proj}{\mathcal{P}}
\newcommand{\qroj}{\mathcal{Q}}
\newcommand{\eigvec}{\ensuremath{\pmb{\epsilon}}}
\newcommand{\Zpart}{\ensuremath{\mathcal{Z}}}
\newcommand{\Vol}{\ensuremath{\mathcal{V}}}
\newcommand{\Cv}{\ensuremath{\mathcal{C}}}
\newcommand{\kBT}{\ensuremath{\mathrm{k}_\mathrm{B}\mathrm{T}}}
\newcommand{\Ra}{\ensuremath{\mathbfcal{R}}}
\newcommand{\bbraket}[1]{\braket{\braket{#1}}}
\newcommand{\bbl}{\ensuremath{\big(}}
\newcommand{\bbr}{\ensuremath{\big)}}
\newcommand{\KLD}{\ensuremath{D_{\mathrm{KL}}}}
\renewcommand\vec{\mathbf}
\newcommand{\phivec}{\boldsymbol{\Phi}}

\preprint{APS/123-QED}

%\title{Thermal conductivity of anharmonic crystals : theory and efficient implementation}
% \title{Temperature dependent generalization of thermal transport in anharmonic crystals : theory and efficient implementation}
\title{Mode-coupling formulation of heat transport in anharmonic materials}

%\author{}

\author{Alo\"is Castellano}
\author{J. P. Alvarinhas Batista}
\affiliation{Nanomat group, Q-MAT research center and European Theoretical Spectroscopy Facility, Université de Liège, allée du 6 août, 19, B-4000 Liège, Belgium}

\author{Olle Hellman}
\affiliation{Department of Molecular Chemistry and Materials Science, Weizmann Institute of Science, Rehovoth 76100, Israel}

\author{Matthieu J. Verstraete}
\affiliation{Nanomat group, Q-MAT research center and European Theoretical Spectroscopy Facility, Université de Liège, allée du 6 août, 19, B-4000 Liège, Belgium}
\affiliation{ITP, Physics Department, Utrecht University 3508 TA Utrecht, The Netherlands}

%\affiliation{%


\begin{abstract}
The temperature-dependent phonons are a generalization of interatomic force constants varying in T, which has found widespread use in computing the thermal transport properties of materials. A formal justification for using this combination to access thermal conductivity in anharmonic crystals, beyond the harmonic approximation and perturbation theory, is still lacking.
    In this work, we derive a theory of heat transport in anharmonic crystals, using the mode-coupling theory of anharmonic lattice dynamics. 
    Starting from the Green-Kubo formula, we develop the thermal conductivity tensor based on the system's dynamical susceptibility, or spectral function. 
    Our results account for both the diagonal and off-diagonal contributions of the heat current, with and without collective effects. 
    We implement our theory in the TDEP package, and have notably introduced a Monte Carlo scheme to compute phonon scattering due to third- and fourth-order interactions, achieving a substantial reduction in computational cost which enables full convergence of such calculations for the first time. 
    We apply our methodology to systems with varying regimes of anharmonicity and thermal conductivity to demonstrate its universality. 
    These applications highlight the importance of the phonon renormalizations and their interactions beyond the harmonic order.
    Overall, our work advances the understanding of thermal conductivity in anharmonic crystals and provides a theoretically robust framework for predicting heat transport in complex materials.
\end{abstract}

\maketitle

%\tableofcontents
\section{Introduction}

Fourier’s law asserts that heat transport is characterized by the thermal conductivity which is an intrinsic material property relating the temperature gradient and heat current. 
This property is essential for selecting candidate materials in many technological applications, each requiring a specific range of thermal conductivity, either high, low, or ``just right''. 
For instance, the development of thermoelectric devices and barrier coatings demands materials with extremely low thermal conductivity~\cite{Liu2019,He2017}. 
Conversely, in applications which generate heat, such as electronic devices, batteries, or nuclear reactors, ensuring safe and controllable operating conditions necessitates the efficient removal of excess heat~\cite{Agne2022,Zeng2021}. 
This can be achieved passively through the high thermal conductivity of the contact and heat sink materials. 
A theoretical understanding of the mechanisms underlying heat transport in materials is thus critical for fundamental science but also many applications.

In electrically insulating solids, the pioneering works of Hardy and Peierls have shown that heat is primarily transported by the vibrations of nuclei around their equilibrium positions. 
Within the harmonic approximation, these vibrations are quantized as quasiparticles called phonons. 
The Peierls-Boltzmann theory describes phonons as heat carriers that diffuse through materials collectively, with their transport limited by scattering due to other quasiparticles, boundaries, or defects~\cite{McGaughey2019,Lindsay2016}. 
Recently, the importance of another transport mechanism has been highlighted, where heat is carried through the wavelike tunneling of phonons to quasidegenerate states. 
This mechanism, derivable from both the Hardy~\cite{Isaeva2019,Srivastava1981,Semwal1972,Hardy1963} and Wigner~\cite{Simoncelli2022,Caldarelli2022} formulations of the heat current, is particularly significant for systems with low thermal conductivity and complex crystal structures. 

Accurately predicting a material’s thermal conductivity thus ultimately comes down to the precise description of atomic vibrations and phonons.
The harmonic approximation relies on a Taylor expansion of the Born-Oppenheimer surface, assuming that displacements around equilibrium positions are relatively small. 
Additionally, perturbation theory is often used to compute scattering mechanisms affecting phonon diffusion, and only converges if higher-order contributions to the potential energy are small compared to the harmonic part. 
However, these assumptions are not always valid in real materials~\cite{Knoop2020,Knoop2023}. This limits the predictive accuracy of the harmonic approximation. 
Recognizing this limitation, theories of temperature-dependent phonons have emerged, notably the self-consistent harmonic approximation~\cite{Koehler1966,Werthamer1970,Tadano2015,Monacelli2021} and the temperature-dependent effective potential~\cite{Hellman2011,Hellman2013a,Hellman2013b}. 
Both approaches involve renormalizing the bare harmonic phonons, through non-perturbative interaction with a bath of all other phonons. 
These methods have been applied to many systems, showing significant improvement over harmonic or perturbative predictions, and highlighting the importance of including anharmonicity in the vibrational description of materials~\cite{Hellman2014,Romero2015,Tadano2018,Fu2022,Wang2023,Yang2022}.

In recent work, we introduced the mode-coupling theory of anharmonic lattice dynamics~\cite{Castellano2023}, providing a formal justification for the temperature-dependent effective potential. 
This theory posits that the phonon bath originates from the full dynamics of the many-body Hamiltonian. 
While applications of this methodology to compute thermal conductivity exist, they are based on formulas derived in a perturbative context, and a formal justification is still lacking.

In this work, we derive a theory of heat transport in anharmonic crystals based on the mode-coupling theory of anharmonic lattice dynamics. 
Our derivation applies heat current operators defined for anharmonic phonons in the Green-Kubo formula, allowing us to construct a theory founded on phonon correlation functions. 
Ultimately, we obtain a formulation of the thermal conductivity tensor that includes anharmonicity non-perturbatively, with both collective and coherent contributions. 
We describe the implementation of the theory in the open-source package TDEP ~\cite{Knoop2024}, emphasizing the reduction of computational cost, which is essential to be able to converge fully with complex unit cells and/or higher order anharmonic coupling. 
Finally, we apply the method to several systems spanning different regimes of anharmonicity and thermal conductivity mechanisms.

The paper is organized as follows.
In section \ref{sec:derivation}, after introducing the mode-coupling theory of anharmonic lattice dynamics, we derive a heat current operator which is consistent with the theory.
This operator is then injected in the Green-Kubo formula, enabling us to obtain formulations of the thermal conductivity tensor.
We then discuss the improvement brought by our approach in \ref{sec:discussion}.
Section \ref{sec:implementation} presents our implementation of the theory in the TDEP package, focusing on approaches to reduce the computational cost.
These methods include a linear algebra formulation of the scattering matrix elements, the irreducible representation of scattering triplets and quartets, and a Monte-Carlo integration scheme for phonon scatterings.
Finally, in section \ref{sec:applications}, we apply our formalism to several materials before concluding in section \ref{sec:conclusion}.

\section{Derivation}
\label{sec:derivation}

We consider a crystalline system within the framework of the Born-Oppenheimer approximation, where the system dynamics is described by the Hamiltonian
\begin{equation}
    H = \sum_i \frac{\vec{P}_i^2}{2 M_i} + V(\vec{R})
\end{equation}
where $\vec{R}$ and $\vec{P}$ are respectively the position and momentum operators and where $V(\vec{R})$ is the many-body potential.
To incorporate quantum effects in the ionic motion, we utilize the quantum Liouvillian formalism, which  describes the time evolution and derivative of an operator $\mathcal{O}$ as
\begin{align}
    \mathcal{O}(t) =& e^{\frac{i}{\hbar}H t} \mathcal{O} e^{-\frac{i}{\hbar}H t} = e^{i\liou t} \mathcal{O} \\
    \dot{\mathcal{O}}(t) =& \frac{i}{\hbar} \big[ H, \mathcal{O}(t) \big] = i\liou \mathcal{O}(t)
\end{align}
where $\liou = [H, \cdot] / \hbar$ is the Liouville superoperator.

We assume that ions oscillate around their equilibrium positions $\braket{\vec{R}}$, allowing us to introduce displacement operators $\vec{u}(t) = \vec{R}(t) - \braket{\vec{R}}$.
In much of the literature, this assumption is used to define an approximate Hamiltonian obtained by truncating a Taylor expansion of the potential energy.
Typically, this expansion is truncated at the third or fourth order, implying small displacement from equilibrium positions.
In our approach, we do not assume a specific amplitude for these displacements, other than ensuring that ions do not diffuse within the crystal, and remain localized around their equilibrium positions.

The focus of this work is the thermal conductivity tensor, which quantifies the heat flux $\mathcal{J}^\alpha$ in Cartesian direction $\alpha$ due to an applied temperature gradient in direction $\beta$
\begin{equation}
    \mathcal{J}^\alpha = - \kappa^{\alpha\beta} \nabla^\beta T
\end{equation}
In the regime where the applied gradient is small, linear response theory dictates that $\kappa$ is an intrinsic equilibrium property expressed by the Green-Kubo formula
\begin{equation}
\label{eq:green kubo}
    \kappa^{\alpha\beta} = \frac{1}{V \kBT^2}\int_0^\infty dt \big( \mathcal{J}^\alpha, \mathcal{J}^\beta(t) \big)
\end{equation}
where $\big(A, B(t)\big) = \kBT \int_0^\beta d\lambda \braket{A(-i\hbar\lambda),B^*(t)}$ denotes the Kubo correlation function (KCF)~\cite{Kubo1966}.
From our initial considerations, the primary challenge lies in expressing heat current operators in terms of the dynamic variables of our systems, specifically the displacements.
However, before addressing this task, it is crucial to establish a comprehensive description of ion dynamics.

\subsection{The mode-coupling theory of anharmonic lattice dynamics}

The many-body nature of the Hamiltonian governing ion motion renders an exact analytical description of the dynamics impossible.
Fortunately, linear response theory offers effective tools for making accurate approximations in such scenarios.
Recently, we introduced the mode-coupling theory of anharmonic lattice dynamics~\cite{Castellano2023}, which uses linear response theory to calculate the correlated motion of ions in crystals beyond standard perturbation theory.
This formalism aims to describe the mass-weighted displacement-displacement KCF~\cite{Kubo1966}
\begin{equation}
\label{eq:KCF}
    G_{ij}^{\alpha\beta}(t) = \sqrt{M_i M_j}\big(u_i^\alpha, u_j^\beta(t) \big)
\end{equation}
In this overview, we outline the key aspects of its derivation and refer interested readers to our previous work for comprehensive details.

The formalism is founded on a Mori-Zwanzig projection scheme~\cite{Mori1965,Zwanzig1961}, with the introduction of the projection operators
\begin{align}
    \proj \cdot =& \sum_{ij\alpha\beta} \frac{\big( u_j^\beta, \cdot\big)}{\big(u_i^\alpha, u_j^\beta\big)} u_i^\alpha + \sum_{ij\alpha\beta} \frac{\big( p_j^\beta, \cdot\big)}{\big(p_i^\alpha, p_j^\beta\big)} p_i^\alpha\\
    \qroj =& 1 - \proj
\end{align}
The operator $\proj$ projects a dynamical variable on the subspace of the full dynamical variables defined by single displacements $u_i$ and their momenta $p_i$, while its orthogonal projection $\qroj$ projects on the rest of the full dynamical variable space.
The first step of the derivation consists in projecting the time derivative of the momentum operator, the atomic force, on both $\proj$ and $\qroj$ which allows, after some steps, to formally write its equation of motion as the generalized Langevin equation~\cite{Kubo1966}
\begin{equation}
\label{eq:forces GLE}
    f_{j}^{\alpha}(t) = -\sum_{k\gamma} \Phi_{jk}^{\beta\gamma} u_{k}^{\gamma}(t) - \sum_{k\gamma}\int_0^t ds K_{jk}^{\beta\gamma}(s) \dot{u}_{k}^{\gamma}(t-s) + \delta f_i(t)
\end{equation}
where we introduced the temperature dependent generalization of the second interatomic force constants (IFC)
\begin{equation}
    \Phi_{ij}^{\alpha\beta} = -\frac{1}{\sqrt{M_i M_j}} \sum_{k\gamma} \frac{\big( u_k^\gamma, f_i^\alpha \big)}{\big( u_j^\beta, u_k^\gamma \big)}
\end{equation}
as well as a memory matrix
\begin{equation}
\label{eq:memory kernel real space}
    K_{ij}^{\alpha\beta}(t) = \beta \frac{\big(\delta f_i^\alpha, \delta f_j^\beta(t) \big)}{\sqrt{M_i M_j}}
\end{equation}
and where $\delta f_i^\alpha(t) = e^{i\qroj \liou t} (f_i^\alpha + \sum_{j\beta} \Phi_{ij}^{\alpha\beta}u_j^\beta)$ is called the ``random'' force due to its projected dynamics outside of the space spanned by single displacements.
Multiplying then by a displacement, taking the Kubo average and using the property $\big(u_i^\alpha, \delta f_j(t)\big)=0$, we obtain the generalized Langevin equation for the equation of motion of the displacement correlation function
\begin{equation}
\label{eq:GLE}
    \ddot{G}_{ij}^{\alpha\beta}(t) = -\sum_{k\gamma} \Phi_{jk}^{\beta\gamma} G_{ik}^{\beta\gamma}(t) - \sum_{k\gamma}\int_0^t ds K_{jk}^{\beta\gamma}(s) \dot{G}_{jk}^{\beta\gamma}(t-s)
\end{equation}

As in the harmonic approximation, the Fourier transform of $\boldsymbol{\Phi}$ allows to define phonons with their associated phonon displacement operator $A_{s}(\vec{q})$, where $s$ is the mode.
The displacements can then be projected onto the phonon space using their eigenvectors $\boldsymbol{\varepsilon}_{s}(\vec{q})$
\begin{equation}
    u_i^\alpha(t) = \sqrt{\frac{\hbar}{2 M_i}} \sum_\lambda \frac{\varepsilon_{s}^{i\alpha}(\vec{q})}{\sqrt{\Omega_{s}(\vec{q})}} A_{s}(\vec{q}, t)
\end{equation}
To simplify the derivation of the heat current and ease the comparison with harmonic and perturbation theory, we will also introduce the phonon momentum operator
\begin{align}
    p_i^\alpha(t) =& -i\sqrt{\frac{\hbar M_i}{2}} \sum_{s\vec{q}} \sqrt{\Omega_s(\vec{q})} \varepsilon_{s}^{i\alpha}(\vec{q}) B_s(\vec{q})
\end{align}
where we can recognize $B_s(\vec{q}, t) = -i \dot{A}_s(\vec{q},t) / \Omega_s(\vec{q})$.
The usefulness of the momentum operator comes from the relation between static correlation function involving $A_s(\vec{q})$ and $B_s(\vec{q})$, for instance
\begin{align}
    \big(A_s(\vec{q}), A_{s'}(\vec{q}')) =& \big(B_s(\vec{q}), B_{s'}(\vec{q}')) = \frac{\kBT}{\hbar \Omega_s(\vec{q})}\delta_{s s'}\delta_{\vec{q} \vec{q}'} \\
    \big(A_s(\vec{q}), B_{s'}(\vec{q}')) =& \big(B_s(\vec{q}), A_{s'}(\vec{q}')) = 0
\end{align}

We can now define the phonon correlation function
\begin{equation}
    G_{s}(\vec{q}, t) = \big(A_{s}(\vec{q}), A_{s}(\vec{q}, t)\big)
    \label{lab:phononKCF}
\end{equation}
which follows the generalized Langevin equation
\begin{equation}
    \ddot{G}_{s}(\vec{q}, t) = -\Omega_{s}^2(\vec{q}) G_{s}(\vec{q}, t) - \int_0^t ds K_{s}(\vec{q},s) \dot{G}_{s}(\vec{q}, t-s)
\end{equation}
where $K_{s}(\vec{q}, t)$ is the projection of the memory kernel on phonon $(\vec{q}, s)$.
Taking the real part of the Laplace transform of this equation, we obtain the phonon correlation function in frequency space
\begin{equation}
\label{eq:green function}
    G_{s}(\vec{q}, \omega) = \frac{\kBT}{\pi\hbar}\frac{8\Omega_{s}(\vec{q}) \Gamma_{s}(\vec{q}, \omega)}{(\omega^2 - \Omega_{s}(\vec{q})^2 - 2 \omega \Delta_{s}(\vec{q}, \omega))^2 + 4 \omega^2 \Gamma_{s}^2(\vec{q}, \omega)}
\end{equation}
where $\Gamma_s(\vec{q}, \omega)$ and $\Delta_{s}(\vec{q}, \omega)$ are the real and imaginary part of the memory kernel, which are related through a Kramers-Kronig transform
\begin{equation}
    \Delta_{s}(\vec{q}, \omega) = \frac{1}{\pi} \int d\omega' \frac{\Gamma_{s}(\vec{q}, \omega')}{\omega' - \omega}
\end{equation}
while $\Gamma_s(\vec{q},\omega)$ is proportional to the Fourier transform of the memory kernel.
It should be noted that up until this point, the only approximation made concerns the neglect in eq.(\ref{lab:phononKCF}) of the off-diagonal component of the correlation function for a given $\vec{q}$-point.
The correlation function in eq.(\ref{eq:green function}) is related to the phonon spectral function, which can be directly compared to experiments such as inelastic neutron or X-ray scattering and is obtained from the fluctuation-dissipation theorem $\chi''_{s}(\vec{q}, \omega) = \frac{\omega}{2\kBT} G_{s}(\vec{q}, \omega)$, resulting in
\begin{equation}
\label{eq:spectral function}
    \chi''_{s}(\vec{q}, \omega) = \frac{1}{\pi\hbar}\frac{4\omega\Omega_{s}(\vec{q}) \Gamma_{s}(\vec{q}, \omega)}{(\omega^2 - \Omega_{s}(\vec{q})^2 - 2 \omega \Delta_{s}(\vec{q}, \omega))^2 + 4 \omega^2 \Gamma_{s}^2(\vec{q}, \omega)}
\end{equation}

\begin{figure}
    \centering
    \includegraphics[width=\columnwidth]{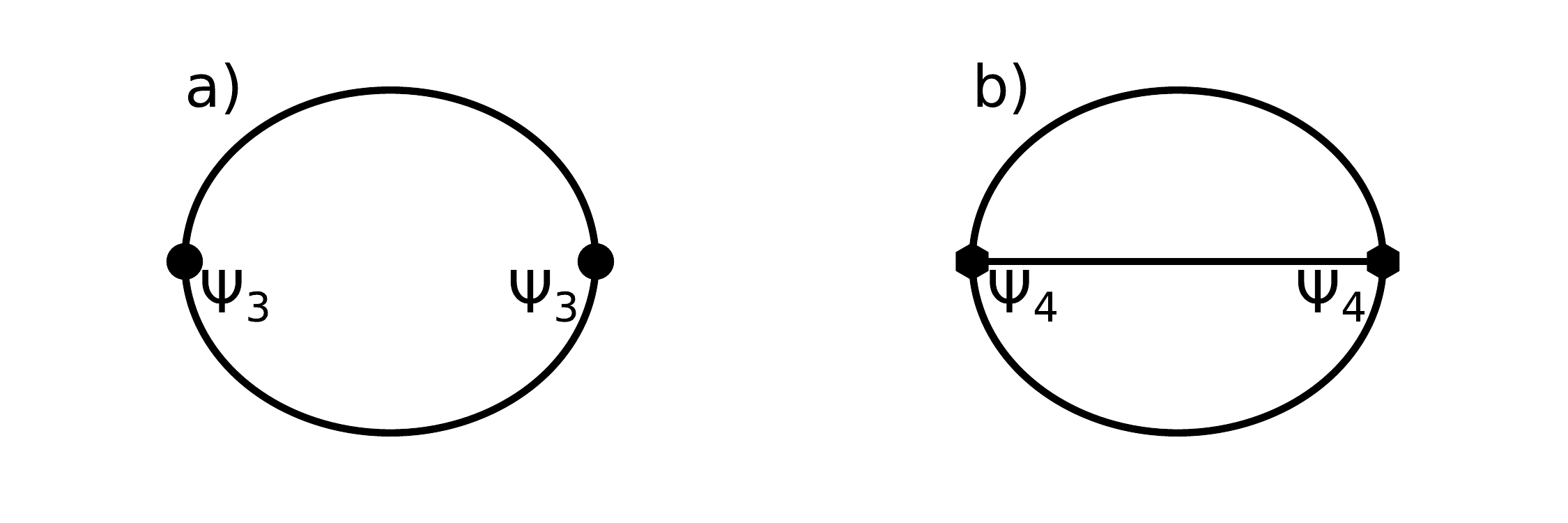}
    \caption{Feynman diagrams for the memory kernel considered in this work.
    a) The three phonon bubble contribution b) the four phonon sunset contribution}
    \label{fig:feynman}
\end{figure}

The main difficulty in employing eq.(\ref{eq:green function}) or (\ref{eq:spectral function}) lies in the a priori unknown expression of the memory kernel.
In the mode-coupling approximation, this difficulty is alleviated by expanding the random forces using higher order displacement projection operators.
Up to fourth order, the random forces are then written as
\begin{equation}
\label{eq:random forces 4th}
\begin{split}
    \delta f_i^\alpha(t) =& e^{i\qroj\liou t} \bigg[\frac{1}{2!}\sum_{jk}\sum_{\beta\gamma}\Psi_{ijk}^{\alpha\beta\gamma} u_j^\beta u_k^\gamma \\
    +& \frac{1}{3!}\sum_{jkl}\sum_{\beta\gamma\delta}\Psi_{ijkl}^{\alpha\beta\gamma\delta} u_j^\beta u_k^\gamma u_l^\delta + \delta_4 f_i^\alpha \bigg]
\end{split}
\end{equation}
where $\delta_4 f_i^\alpha$ is the remainder of the force and the $\boldsymbol{\Psi}$ are the temperature-dependent generalizations of higher-order force constants.
These can be computed as
\begin{align}
    \Psi_{ijk}^{\alpha\beta\gamma} =& \sum_{k'l'}\sum_{\beta'\gamma'} \frac{\big( u_{j'}^{\beta'}u_{k'}^{\gamma'}, \delta f_i^\alpha \big)}{\big( u_{j'}^{\beta'}u_{k'}^{\gamma'}, u_{j}^\beta u_{k}^\gamma \big)} \\
    \Psi_{ijkl}^{\alpha\beta\gamma\delta} =& \sum_{j'k'l'}\sum_{\beta'\gamma'\delta'} \frac{\big( u_{j'}^{\beta'}u_{k'}^{\gamma'}u_{l'}^{\delta'}, \delta_3f_i^\alpha\big)}{\big( u_{j'}^{\beta'}u_{k'}^{\gamma'}u_{l'}^{\delta'}, u_{j}^\beta u_{k}^\gamma u_{l}^\delta \big)}
\end{align}
where $\delta_3 f_i^\alpha = \delta f_i^\alpha - \sum_{jk}\sum_{\beta\gamma} \Psi_{ijk}^{\alpha\beta\gamma} u_j^\beta u_k^\gamma$.
An approximation up to fourth order of the memory matrix can be computed by injecting eq.(\ref{eq:random forces 4th}) in eq.(\ref{eq:memory kernel real space}) after the neglect of the $\delta_4 f_i^\alpha$ term and of the orthogonal projector $\qroj$ in the time evolution.
After a projection on phonon modes, the memory kernel for mode $(s,\vec{q})$ can be decomposed as
\begin{equation}
    \Gamma_{s}(\vec{q}, \omega) \approx \Gamma_{s}^{(3)}(\vec{q}, \omega) + \Gamma_{s}^{(4)}(\vec{q}, \omega)
\end{equation}
After decoupling the various correlation functions appearing in $\Gamma_s(\vec{q})$ using the scheme presented in appendix~\ref{app:decoupling}, one obtains a set of self-consistent equations for the memory kernel and $G_s(\vec{q}, \omega)$.
This set can be replaced by a one-shot approximation, where the phonon correlation functions involved in the memory kernel are replaced by their memory-free counterparts.
%\begin{widetext}
In this approximation, the three phonon contribution is written
\begin{widetext}
\begin{align}
    \Gamma_{s_1}^{(3)}(\vec{q}_1, \omega) =&  \sum_{\vec{q}_2\vec{q}_3} \sum_{s_2 s_3} \Gamma_{s_1 s_2 s_3}^{(3)}(\vec{q}_1, \vec{q}_2, \vec{q}_3, \omega) \\
    \Gamma_{s_1 s_2 s_3}^{(3)}(\vec{q}_1, \vec{q}_2, \vec{q}_3, \omega) =& \frac{\pi}{16} \vert \Psi_{\vec{q}_1\vec{q}_2\vec{q}_3}^{s_1 s_2 s_3} \vert^2 \mathcal{S}^{(3)}(\omega, \Omega_{s_2}({\vec{q}_2}), \Omega_{s_3}({\vec{q}_3})) \\
    \mathcal{S}^{(3)}(\omega, \Omega_2, \Omega_3) =& \sum_{s=1,-1} \big[s \left( n(\Omega_2) + n(\Omega_3) + 1\right) \delta(\omega + s\Omega_2 + s\Omega_3) +
   s\left(n(\Omega_2) - n(\Omega_3)) \delta(\omega + s\Omega_2 - s\Omega_3\right)\big]
\end{align}
while the four phonon interaction is given by
\begin{align}
    \Gamma_{s_1}^{(4)}(\vec{q}_1, \omega) =& \sum_{\vec{q}_2\vec{q}_3 \vec{q}_4} \sum_{s_2 s_3 s_4} \Gamma_{s_1 s_2 s_3 s_4}^{(4)}(\vec{q}_1, \vec{q}_2, \vec{q}_3, \vec{q}_4, \omega)\\
    \Gamma_{s_1 s_2 s_3 s_4}^{(4)}(\vec{q}_1, \vec{q}_2, \vec{q}_3, \vec{q}_4, \omega) =& \frac{\pi}{96} \vert \Psi_{\vec{q}_1\vec{q}_2\vec{q}_3\vec{q}_4}^{s_1 s_2 s_3 s_4} \vert^2 \mathcal{S}^{(4)}(\omega, \Omega_{s_2}({\vec{q}_2}), \Omega_{s_3}({\vec{q}_3}), \Omega_{s_4}({\vec{q}_4})) \\
    \mathcal{S}^{(4)}(\omega, \Omega_2, \Omega_3, \Omega_4) =& \sum_{s=1,-1} \big[ s\left(n(\Omega_2) + 1) (n(\Omega_3) + 1)( n(\Omega_4) + 1) - n(\Omega_2) n(\Omega_3) n(\Omega_4)\right) \delta(\omega + s\Omega_2 + s\Omega_3 + s\Omega_4)  \\
    &+ s\left( 3 n(\Omega_2) (n(\Omega_3) + 1) (n(\Omega_4) + 1) - (n(\Omega_2) + 1) n(\Omega_3) n(\Omega_4) \right) \delta(\omega + s\Omega_2 - s\Omega_3 - s\Omega_4)\big]
\end{align}
\end{widetext}
In these equations, the scattering matrix elements $\Psi_{\vec{q}_1\vec{q}_2 \vec{q}_3}^{s_1 s_2 s_3}$ and $\Psi_{\vec{q}_1\vec{q}_2\vec{q}_3\vec{q}_4}^{s_1 s_2 s_3 s_4}$ are the projections of the higher order generalized IFCs on phonon modes.
Introducing a unit cell centered notation $\Psi_{ijk}^{\alpha\beta\gamma}(\mu,\nu)$, where $i$, $j$ and $k$ are atoms in the unit cell and $\mu$ and $\nu$ denotes the index of a unit cell in the crystal, the third order scattering matrix elements are computed as
\begin{align}
    \begin{split}
    \label{eq:scatmat old}
    \Psi_{\vec{q}_1\vec{q}_2 \vec{q}_3}^{s_1 s_2 s_3} = \sum_{ijk \alpha\beta\gamma\mu\nu}& \frac{\Psi_{ijk}^{\alpha\beta\gamma}(\mu,\nu) e^{-i(\vec{q}_2 \vec{R}_{\mu,j} + \vec{q}_3 \vec{R}_{\nu,k})}}{\sqrt{M_i M_j M_k}} \\
    &\frac{\boldsymbol{\varepsilon}_{s_1}^{i\alpha}(\vec{q}_1) \boldsymbol{\varepsilon}_{s_2}^{j\beta}(\vec{q}_2) \boldsymbol{\varepsilon}_{s_3}^{k\gamma}(\vec{q}_3)}{\sqrt{\Omega_\lambda \Omega_{\lambda'} \Omega_{\lambda''}}} \\
    &\Delta(\vec{q}_1 + \vec{q}_2 + \vec{q}_3)
    \end{split} %\\
%   \begin{split}
%   \Psi_{\vec{q}_1\vec{q}_2\vec{q}_3\vec{q}_4}^{s_1 s_2 s_3 s_4} = \sum_{ijkl\alpha\beta\gamma\delta}& \frac{\Psi_{ijkl}^{\alpha\beta\gamma\delta}(\mu,\nu,\lambda) e^{-i(\vec{q}_2 \vec{R}_{\mu,j} + \vec{q}_3 \vec{R}_{\nu,k} + \vec{q}_4\vec{R}_{\lambda,l})}}{\sqrt{M_i M_j M_k M_l}} \\
%   &\frac{\boldsymbol{\varepsilon}_{s_1}^{i\alpha}(\vec{q}_1) \boldsymbol{\varepsilon}_{s_2}^{j\beta}(\vec{q}_2) \boldsymbol{\varepsilon}_{s_3}^{k\gamma}(\vec{q}_3) \boldsymbol{\varepsilon}_{s_4}^{l\delta}(\vec{q}_4)}{\sqrt{\Omega_{s_1}(\vec{q}_1) \Omega_{s_2}(\vec{q}_2) \Omega_{s_3}(\vec{q}_3) \Omega_{s_4}(\vec{q}_4)}} \\
%   &\Delta(\vec{q}_1 + \vec{q}_2 + \vec{q}_3 + \vec{q}_4)
%   \end{split}
\end{align}
%   \begin{align}
%       \begin{split}
%       \label{eq:scatmat old}
%       \Psi_{\vec{q}_1\vec{q}_2 \vec{q}_3}^{s_1 s_2 s_3} = \sum_{ijk \alpha\beta\gamma}& \frac{\Psi_{ijk}^{\alpha\beta\gamma} e^{-i(\vec{q}_2 \vec{R}_j + \vec{q}_3 \vec{R}_k)}}{\sqrt{M_i M_j M_k}} \\
%       &\frac{\boldsymbol{\varepsilon}_{s_1}^{i\alpha}(\vec{q}_1) \boldsymbol{\varepsilon}_{s_2}^{j\beta}(\vec{q}_2) \boldsymbol{\varepsilon}_{s_3}^{k\gamma}(\vec{q}_3)}{\sqrt{\Omega_\lambda \Omega_{\lambda'} \Omega_{\lambda''}}} \\
%       &\Delta(\vec{q}_1 + \vec{q}_2 + \vec{q}_3)
%       \end{split} \\
%       \begin{split}
%       \Psi_{\vec{q}_1\vec{q}_2\vec{q}_3\vec{q}_4}^{s_1 s_2 s_3 s_4} = \sum_{ijkl\alpha\beta\gamma\delta}& \frac{\Psi_{ijkl}^{\alpha\beta\gamma\delta} e^{-i(\vec{q}_2 \vec{R}_j + \vec{q}_3 \vec{R}_k + \vec{q}_4\vec{R}_l)}}{\sqrt{M_i M_j M_k M_l}} \\
%       &\frac{\boldsymbol{\varepsilon}_{s_1}^{i\alpha}(\vec{q}_1) \boldsymbol{\varepsilon}_{s_2}^{j\beta}(\vec{q}_2) \boldsymbol{\varepsilon}_{s_3}^{k\gamma}(\vec{q}_3) \boldsymbol{\varepsilon}_{s_4}^{l\delta}(\vec{q}_4)}{\sqrt{\Omega_{s_1}(\vec{q}_1) \Omega_{s_2}(\vec{q}_2) \Omega_{s_3}(\vec{q}_3) \Omega_{s_4}(\vec{q}_4)}} \\
%       &\Delta(\vec{q}_1 + \vec{q}_2 + \vec{q}_3 + \vec{q}_4)
%       \end{split}
%   \end{align}
with $\vec{R}_{\mu,j}$ the distance between the atom $i$ in a reference unitcell and the atom $j$ in the unit cell $\mu$ and where $\Delta(\vec{q})$ is $1$ if $\vec{q}$ is equal to a reciprocal lattice vector and $0$ otherwise, to ensure conservation of the quasi-momentum.
The fourth-order scattering matrix elements are computed with a similar formula involving the fourth order generalized IFC.
The memory kernel in the mode-coupling approximation can be rationalized from a diagrammatic representation, pictured in Fig.(\ref{fig:feynman}). 
In this representation, the third order contribution is analogous to the third order bubble diagram from perturbation theory, while the fourth order is equivalent to the sunset diagram.

It should be noted that to introduce the scattering of phonons due to isotopic effects, the following contribution can be added to the memory kernel
\begin{equation}
\begin{split}
 \Gamma_{s}^{\mathrm{iso}}(\vec{q}, \omega) =& \sum_{\vec{q}_2 s_2} \Gamma_{s s_2}^{\mathrm{iso}}(\vec{q}, \vec{q}_2, \omega) \\
 =& \sum_{\vec{q}_2 s_2} \sum_i g_i \vert \boldsymbol {\varepsilon}_s^i(\vec{q}) {\varepsilon}_{s_2}^i(\vec{q}_2) \vert^2 \delta(\omega - \Omega_{s_2}(\vec{q}_2))
\end{split}
\end{equation}
In this equation, corresponding to Tamura's model~\cite{Tamura1983}, the $g_i$ measure the distribution of the isotope masses of element $i$ and is computed as $g_i=\sum_n \frac{d_{i,n}}{N} \bigg( \frac{\Delta M_{i,n}}{M_{i}} \bigg)^2$ where $N$ is the number of isotopes, $d_{i,n}$ is the concentration of isotope $n$ of element $i$, and $\Delta M_{i,n}$ is the mass difference between isotope $n$ and the average mass of the element.

The generalized Langevin equation of eq.(\ref{eq:GLE}) describes phonons interacting with a bath made from all other phonons.
When the interaction with this bath is weak, the time dependence of the memory kernel can be neglected, and the dissipative part of the dynamics can be reduced to a single value $\Gamma_s^M(\vec{q})$, given by $\Gamma_s(\vec{q}, \omega)$ evaluated at the frequency of the phonon $(s,\vec{q})$, as shown in appendix~\ref{app:markovian}
\begin{equation}
    \Gamma_s^M(\vec{q}) = \Gamma_s(\vec{q}, \Omega_s(\vec{q}))
\end{equation}
In this limit, known as Markovian, the spectral function reduces to a Lorentzian centered at the frequency $\Omega_s(\vec{q})$ and with a width $\Gamma_s(\vec{q})$
\begin{equation}
    \chi''^M_s(\vec{q}, \omega) = \frac{1}{\pi \hbar}\frac{\Gamma_s^M(\vec{q})}{(\omega - \Omega_s(\vec{q}))^2 + \big[\Gamma_s^M(\vec{q})\big]^2}
\end{equation}
The Markovian limit cannot be a complete description of the system, since it breaks some sum rules that the correlation functions are supposed to follow~\cite{Castellano2023}. Nevertheless, it remains a useful and often accurate approximation for systems where the quasiparticle picture is well founded.
In this case, the phonons can be thought of as diffusing through the material with a lifetime $\tau_s(\vec{q}) = \big[2\Gamma_s^M(\vec{q})\big]^{-1}$.
One can note that the Markovian limit is analogous to the use of Fermi's golden rule in perturbation theory.

\subsection{The anharmonic heat current operator}

To derive the thermal conductivity tensor using the Green-Kubo formula, it is essential to derive heat current operators which are consistent with the previously established mode-coupling theory.
In an electrically insulating solid, with our prerequisite of absence of diffusion, the conductive component of the heat current operator, defined as~\cite{Marcolongo2015,Isaeva2019}
\begin{equation}
    \boldsymbol{\mathcal{J}}(t) = \sum_i \braket{\vec{R}_i} \dot{E}_i(t)
\end{equation}
is the only contributor the thermal conductivity.
Here, $E_i(t)$ represents the local energy contribution from atom $i$ to the system's total energy, a quantity that is inherently ambiguous.
To circumvent the complexities associated with explicitly partitioning the potential energy, we instead focus directly on the time derivative $\dot{E}_i(t)$.
Building on the approach of \cite{Carbogno2017}, the heat current operator can be expressed as
\begin{equation}
    \boldsymbol{\mathcal{J}}(t) = \frac{1}{V}\sum_{ij} (\braket{\vec{R}_i} - \braket{\vec{R}_j})\vec{f}_{ij}(t) \dot{\vec{u}}_i(t)
\end{equation}
where $\vec{f}_{ij}(t)$ denotes the forces exerted by atom $j$ on atom $i$.
To formally define this quantity, one would typically require an explicit formulation of the Hamiltonian.
However, the Mori-Zwanzig projection scheme offers an alternative approach to perform the partitioning.
%Using the equation of motion derived in \cite{Castellano2023} and imposing $\vec{f}_i(t) = \sum_{j} \vec{f}_{ij}(t)$, we arrive at the partitioned form
Indeed, from the equation of motion for the forces (eq.(\ref{eq:forces GLE})), we can  partition the force over atom pairs, resulting in the expression
\begin{equation}
    \vec{f}_{ij}^\alpha(t) = -\Phi_{ij}^{\alpha\beta} u_j^\beta(t) - \int_0^\infty ds K_{ij}^{\alpha\beta}(s) p_j^\beta(t-s) + \delta f_j^\alpha(t)\delta_{ij}
\end{equation}
which respects the condition $\vec{f}_i(t) = \sum_j \vec{f}_{ij}(t)$.
Neglecting the contribution from the memory kernel and the random force, we can substitute this expression into the heat current operator, which yields
\begin{equation}
    \mathcal{J}^\alpha(t) \approx \sum_{ij}\sum_{\beta\gamma} (\braket{R_i^\alpha} - \braket{R_j^\alpha}) \Phi_{ij}^{\beta\gamma} u_j^\gamma(t) \dot{u}_i^\beta(t)
\end{equation}
which, in term of phonons operator, becomes
\begin{equation}
\label{eq:heat current end}
    \boldsymbol{\mathcal{J}}(t) = -\frac{1}{2} \sum_{\vec{q} s_1 s_2} \hbar \Omega_{s_2}(\vec{q}) \vec{v}_{\vec{q}}^{s_1 s_2} A_{s_1}(\vec{q}, t) B_{s_2}(\vec{q}, t)
\end{equation}
with $\vec{v}_{\vec{q}}^{s_1 s_2}$ the generalized group velocities~\cite{Isaeva2019,Dangi2021,Caldarelli2022}.
The heat current can be split as $\mathcal{J}(t) = \mathcal{J}^\mathrm{d}(t) + \mathcal{J}^\mathrm{nd}(t)$, where the first term is diagonal with respect to phonon branches
\begin{equation}
\label{eq:heat current diag}
    \boldsymbol{\mathcal{J}}^{\mathrm{d}}(t) = -\frac{1}{2} \sum_{\vec{q} s} \hbar \Omega_{s}(\vec{q}) \vec{v}_{\vec{q}}^{s} A_{s}(\vec{q}, t) B_{s}(\vec{q}, t)
\end{equation}
with $\vec{v}_\vec{q}^s = \vec{v}_{\vec{q}}^{s s}$, and the other term is the off-diagonal contribution
\begin{equation}
\label{eq:heat current nd}
    \boldsymbol{\mathcal{J}}^{\mathrm{nd}}(t) = -\frac{1}{2} \sum_{\vec{q} s_1 \neq s_2} \hbar \Omega_{s_2}(\vec{q}) \vec{v}_{\vec{q}}^{s_1 s_2} A_{s_1}(\vec{q}, t) B_{s_2}(\vec{q}, t)
\end{equation}
Neglecting the correlations between diagonal and off diagonal heat current, the thermal conductivity can be separated in a similar manner
\begin{align}
    \boldsymbol{\kappa} \approx& \boldsymbol{\kappa}^{\mathrm{d}} + \boldsymbol{\kappa}^{\mathrm{nd}} \\
    \boldsymbol{\kappa}^{\mathrm{d}} =& \frac{1}{VT} \int_0^\infty \big( \mathcal{J}^{\mathrm{d}}, \mathcal{J}^{\mathrm{d}}(t) \big) \\
    \boldsymbol{\kappa}^{\mathrm{nd}} =& \frac{1}{VT} \int_0^\infty \big( \mathcal{J}^{\mathrm{nd}}, \mathcal{J}^{\mathrm{nd}}(t) \big)
\end{align}

\subsection{The diagonal thermal conductivity}

We will begin with the diagonal part of $\boldsymbol{\kappa}$, 
\begin{equation}
\label{eq:kappa diag 1}
\begin{split}
    \boldsymbol{\kappa}^{\mathrm{d}} =& \frac{1}{4V\kBT^2} \sum_{\vec{q}_1\vec{q}_2} \sum_{s_1 s_2} \Omega_{s_1}(\vec{q}_1) \Omega_{s_2}(\vec{q}_2) \vec{v}_{\vec{q}_1}^{s_1} \otimes \vec{v}_{\vec{q}_2}^{s_2} \\
    & \int_0^\infty dt \big( A_{s_1}(\vec{q}_1)B_{s_1}(\vec{q}_1), A_{s_2}(\vec{q}_2, t)B_{s_2}(\vec{q}_2, t)\big)
\end{split}
\end{equation}
The main difficulty in this equation lies in the expression of the four point correlation function.
In appendix \ref{app:scattering matrix}, we show how to obtain its equation of motion in the mode-coupling theory, which, after a Laplace transform and the application of the Markovian approximation, allows to express the diagonal contribution to $\boldsymbol{\kappa}$ as
\begin{equation}
\label{eq:kappa diag bte}
    \boldsymbol{\kappa}^{\mathrm{d}} = \frac{1}{V}\sum_{\vec{q}_1\vec{q}_2}\sum_{s_1 s_2} \vec{v}_{\vec{q}_1}^{s_1}\otimes\vec{v}_{\vec{q}_2}^{s_2} c_{s_1}(\vec{q}_1) (\boldsymbol{\Xi}^{-1})(\vec{q}_1 s_1, \vec{q}_2 s_2)
\end{equation}
where $c_s(\vec{q}) = \Omega_s^2(\vec{q}) n_s(\vec{q}) (n_s(\vec{q}) + 1) / \kBT^2$ is the modal heat capacity and $\boldsymbol{\Xi}$ is the scattering matrix, given explicitly in appendix \ref{app:scattering matrix}.

Neglecting the off-diagonal component of the scattering matrix, we obtain the single mode approximation to the thermal conductivity tensor
\begin{equation}
    \boldsymbol{\kappa}^{\mathrm{d,SMA}} = \frac{1}{V} \sum_{\vec{q} s} \vec{v}_\vec{q}^s\otimes\vec{v}_\vec{q}^s \frac{c_s(\vec{q})}{2 \Gamma_s^M(\vec{q})}
\end{equation}
It is interesting to note that this result can also be obtained using the decoupling scheme of Kubo correlation functions, as we will use for the non-diagonal contribution.

\subsection{The non-diagonal contribution to the thermal conductivity}

The contribution stemming from the off-diagonal part of the heat current is written
\begin{equation}
\label{eq:start kappa offdiag}
\begin{split}
    \boldsymbol{\kappa}^{\mathrm{d}} =& \frac{1}{4V\kBT^2} \sum_{\vec{q}_1\vec{q}_2} \sum_{s_1 s_2}' \sum_{s_3 s_4}' \Omega_{s_2}(\vec{q}_1) \Omega_{s_4}(\vec{q}_2) \vec{v}_{\vec{q}_1}^{s_1 s_2} \otimes \vec{v}_{\vec{q}_2}^{s_3 s_4} \\
    & \int_0^\infty dt \big( A_{s_1}(\vec{q}_1)B_{s_2}(\vec{q}_1), A_{s_3}(\vec{q}_2, t)B_{s_4}(\vec{q}_2, t)\big)
\end{split}
\end{equation}
where $\sum_{s_1 s_2}'$ indicates that $s_1 = s_2$ is excluded from the double sum.
Using the rules presented in appendix \ref{app:decoupling}, the four-point Kubo correlation function can be decoupled as
\begin{equation}
\begin{split}
    & \int_0^\infty dt \big( A_{s_1}(\vec{q}_1)B_{s_2}(\vec{q}_1), A_{s_3}(\vec{q}_2, t)B_{s_4}(\vec{q}_2, t)\big) \\
    &\approx \delta_{\vec{q}_1 \vec{q}_2}\delta_{s_1 s_3}\delta_{s_2 s_4}\frac{\pi}{\Omega_{s_1}(\vec{q}_1)\Omega_{s_2}(\vec{q}_1)} \times \\
    &\int_{-\infty}^{\infty} d\omega  \chi_{s_1}''(\vec{q}, \omega)\chi_{s_2}''(\vec{q}, \omega) \omega^2 n(\omega) (n(\omega) + 1)
\end{split}
\end{equation}
Injecting this result in eq.(\ref{eq:start kappa offdiag}), and introducing the frequency dependent heat capacity $c_v(\omega) = \hbar\omega^2 n(\omega) (n(\omega) + 1)/\kBT^2$, the off-diagonal contribution to the thermal conductivity tensor is
\begin{equation}
\label{eq:off diagonal nonmarkov}
    \begin{split}
    \boldsymbol{\kappa}^{\mathrm{nd}} \approx& \frac{\pi}{V} \sum_{\vec{q}}\sum_{s_1 s_2}' \vec{v}_{\vec{q}}^{s_1 s_2} \otimes \vec{v}_{\vec{q}}^{s_1 s_2} \\
    &\int_{-\infty}^\infty d\omega \chi''_{s_1}(\vec{q}, \omega) \chi''_{s_2}(\vec{q}, \omega) c_v(\omega)
    \end{split}
\end{equation}
Within the Markovian approximation, this equation involves the integral of 2 Lorentzians multiplied by the heat capacity of a harmonic oscillator.
If we assume a regime of anharmonicity where the quasiparticle picture is valid, each Lorentzian approaches a Dirac delta, allowing us to take the approximation
\begin{equation}
\label{eq:off diagonal markov}
    \boldsymbol{\kappa}^{\mathrm{nd}} \approx \frac{1}{V} \sum_{\vec{q}} \sum_{s_1 s_2}' \vec{v}_{\vec{q}}^{s_1 s_2} \otimes \vec{v}_{\vec{q}}^{s_1 s_2} \frac{c_{\vec{q}}^{s_1} + c_\vec{q}^{s_2}}{2} \Gamma_{s_1 s_2}^M(\vec{q})
\end{equation}
    where we introduced the off diagonal scattering
\begin{equation}
    \Gamma_{s_1 s_2}^M(\vec{q}) = \frac{\Gamma_{s_1}^M(\vec{q}) + \Gamma_{s_2}^M(\vec{q})}{(\Omega_{s_1}(\vec{q}) - \Omega_{s_2}(\vec{q}))^2 + (\Gamma_{s_1}^M(\vec{q}) + \Gamma_{s_2}^M(\vec{q}))^2}
\end{equation}
One should note that by removing the restriction of $s_1$ and $s_2$ being different, the diagonal contribution in the single mode approximation is recovered, as expected.

\section{Discussion}
\label{sec:discussion}

Our final formulation of the thermal conductivity tensor is expressed as $\boldsymbol{\kappa} = \boldsymbol{\kappa}^{\mathrm{d}} + \boldsymbol{\kappa}^{\mathrm{nd}}$, with $\boldsymbol{\kappa}^{\mathrm{d}}$ given by Eq. (\ref{eq:kappa diag bte}) and $\boldsymbol{\kappa}^{\mathrm{nd}}$ by Eq. (\ref{eq:off diagonal markov}).
This result bears some resemblance to previous derivations based on Hardy’s formulation of the heat current operator \cite{Hardy1963, Isaeva2019, Caldarelli2022, Han2022, Fiorentino2023, Semwal1972, Srivastava1981}.
For instance, we show in appendix \ref{app:pbe} that the diagonal contribution of the thermal conductivity tensor $\boldsymbol{\kappa}^{\mathrm{d}}$ is equivalent to that obtained through the solution of the phonon Boltzmann equation, providing a further proof with reference~\cite{Fiorentino2023} of the equivalence between the phonon Boltzmann and Green Kubo approaches.
Thus, our derivation provides a unified framework that encompasses the collective, single-mode and off-diagonal contributions to heat transport by phonons.
Notably, for the single-mode and off-diagonal contributions, it captures the non-Markovian memory effects described in eq.(\ref{eq:off diagonal nonmarkov}) through the inclusion of the full phonon dynamical susceptibility.
As a result, our formulation can address any system with a crystalline reference structure, from highly harmonic crystals at low temperatures—where collective effects dominate—to complex crystals with large unit cells, where the off-diagonal components of $\boldsymbol{\kappa}^{\mathrm{nd}}$ are essential

A key distinction of our approach is its inherent inclusion of temperature-dependent phonon renormalization, setting it apart from standard formalisms.
The latter rely on a Taylor expansion of the Born-Oppenheimer potential energy surface, treating anharmonic terms as a perturbative correction to a dominant second-order term.
Consequently, the dynamical properties of the system are inferred indirectly, being reconstructed \textit{a posteriori} from the effective potential energy surface.
However, in order to use perturbation theory, these methods assume that the atoms vibrate closely around their equilibrium positions, an assumption that fails at elevated temperatures, in the presence of nuclear quantum effects, or when the Hessian of the Born-Oppenheimer surface is not positive definite.
In these scenarios, high order anharmonic interactions become significant, making perturbative corrections insufficient.

In contrast, our mode-coupling theory focuses directly on the atom dynamics, rather than on the underlying potential energy surface.
Because of this, the framework naturally incorporates temperature-dependent interactions and avoids the limitations of perturbative expansions.

In the end, this difference in foundations is critical.
To understand how the approaches diverge, it is useful to focus on their fundamental building blocks: the interatomic force constants and the phonons.
In the harmonic case (and its perturbation expansion), the IFC are derivatives of the Born-Oppenheimer surface.
At the second order, the effective Hamiltonian can be diagonalized, giving rise to eigenstates: the harmonic phonons.
Interactions between these phonons are then introduced through the higher order IFCs, with the magnitude of phonon-phonon interactions at a given temperature being proportional to these IFCs and to the phonon population.

However, in general, these harmonic phonons are a very rough approximation of the true dynamical quantities.
Indeed, since the second-order term captures only part of the full potential energy landscape, the accuracy of the harmonic phonons as descriptors of atomic motion is inherently limited.
For instance, it has been shown that, in some systems, the harmonic component can account for less than half of the forces acting on atoms~\cite{Knoop2020}.
In such cases, the validity of perturbation theory is compromised: not only does the non-interacting phonon baseline inadequately describe the dynamics, but the phonon-phonon interactions themselves are poorly captured and constrained by the finite order of the Taylor expansion.

In contrast, the mode-coupling theory is constructed to alleviate these shortcomings.
For example, we have shown previously~\cite{Castellano2023} that the second-order generalized IFCs are physically meaningful, being proportional to the inverse static susceptibility.
As a result, in the static limit, the phonons defined by mode-coupling theory are exact, corresponding to the mass-weighted displacement covariance.
Furthermore, the Mori-Zwanzig projection scheme ensures that these phonons provide a minimally interacting basis, representing the most accurate non-interacting phonons possible, with already exact static properties~\cite{Castellano2023}.
On top of this, each order of the mode-coupling approximation is built to minimize the amplitude of all subsequent orders.

Thus, mode-coupling theory provides a rigorous and systematic framework for capturing dynamical properties.
For the thermal conductivity tensor, this approach introduces two primary improvements.
First, it enhances the accuracy of key parameters of the heat current (eq. (\ref{eq:heat current end})), specifically the phonon frequencies and group velocities, due to the exactness of the second-order generalized IFCs.
As a result, both the propagation (through $\vec{v}_\vec{q}^s$) and the amount of heat carried for each phonon mode (through $c_s(\vec{q})$) are more accurately represented.
Second, the mode-coupling theory offers a refined treatment of phonon-phonon scattering, yielding a more precise dynamical description and leading to improved predictions of thermal conductivity.
However, our work is a further demonstration that, even with the renormalization, heat transport is not mediated by individual phonons but is a collective effect, and reduces to phonons only if the coupling between phonons is negligible.
Nevertheless, the refinement brought by the mode-coupling theory also holds in systems where these collective excitations (also known as  relaxons~\cite{Cepellotti2016}), are necessary for an accurate description.
In this case, the equivalence between our derivation and the phonon Boltzmann equation shown in appendix~\ref{app:pbe} further suggests that diagonalizing $\boldsymbol{\Xi}$ allows to obtain the temperature-renormalized counterpart of the relaxons from a perturbative approach.

A notable strength of the mode-coupling theory is that despite its dynamical foundation, its building blocks are real-space and time-independent properties.
Specifically, the generalized IFCs are derived from static Kubo averages, offering distinct advantages over fully time-dependent approaches.

First, this formulation facilitates the evaluation of thermodynamic and long-range limits of correlation functions using relatively moderate simulation sizes and durations.
In contrast, direct time-dependent methods often require significantly larger and more computationally intensive simulations to achieve convergence, making the mode-coupling approach both more efficient and less prone to size-related artifacts.

Additionally, nuclear quantum effects are naturally integrated within the mode-coupling framework, as the formalism is rooted in Kubo correlation functions.
These quantum effects can be explicitly incorporated in practice through path-integral simulations to compute the generalized IFCs.
This is a noticeable advantage, since the path-integral molecular dynamics formalism is only exact in the static limit~\cite{Craig2004} and provides an approximation of the real Kubo correlation that can be spoiled by numerical artifacts such as spurious resonances or shifts in frequency resolved spectra~\cite{Rossi2014,Shiga2008,Witt2009}.
Finally, the the framework being grounded on Kubo correlation functions, supports a rigorously justified semi-classical approximation~\cite{Castellano2023,Ramirez2004}.
By using classical simulations to compute the generalized IFCs, these quantities can then serve as inputs for the quantum equations of motion developed in this work, enabling a practical treatment of quantum nuclear effects.

\section{Implementation}
\label{sec:implementation}

The formalism derived above has been implemented in the TDEP code~\cite{Knoop2024} and this section outlines the strategy used in this implementation. For high-order many body calculations, the latter is not just a question of efficiency: it is crucial to obtain converged results at all. This has been an important and unrecognized problem in comparing different approaches in the literature.

The generalized IFC are fit using linear least-squares on the forces, incorporating the symmetry reduction described in reference~\cite{Hellman2013a}.
This ensures strict adherence to transposition and point-group symmetries, as well as the acoustic and rotational sum rules and Hermiticity~\cite{Maradudin1968}.
A key point of the implementation is the successive fitting of the IFC, meaning that each order is fit on the residual forces from the previous order.
As demonstrated previously~\cite{Castellano2023}, this method aligns with the definition of the generalized IFC in the mode-coupling theory, providing a crucial step beyond the harmonic approximation and perturbation theory.
It should be noted that for systems exhibiting significant nuclear quantum effects, path-integral molecular dynamics can be used, the static KCF needed to compute the generalized IFC corresponding to correlations of the centroid of the quantum polymer.
%For systems exhibiting significant nuclear quantum effects, the centroid of the quantum polymer should be used as a representation of the KCF appearing in the mode-coupling equations.

At the beginning of the thermal conductivity computation, harmonic properties (frequencies, eigenvectors, and group velocities) are generated on a \textbf{q}-point grid.
At this point, the degeneracies are carefully treated following the process introduced in appendix \ref{app:degen}.
To ensure the conservation of the quasi-momentum in the definition of the scattering matrix elements, we use regular grids of size $N_1\times N_2 \times N_3$ where \textbf{q}-points are defined as $\vec{q}=(i_1/ a, i_2 / b, i_3 / c)$, $i_x$ being integers from 0 to $N_x-1$, and $a$, $b$, and $c$ representing the lattice constants of the system~\cite{Li2014,Togo2023}.
This grid structure ensures that given two \textbf{q}-points $\vec{q}_1$ and $\vec{q}_2$, it is always possible to find a third \textbf{q}-point $\vec{q}_3$ in the grid, such that $\vert \vec{q}_3 \vert = \vert \vec{q}_1 + \vec{q}_2 \vert$, thereby enforcing the quasi-momentum conservation.
The same principle applies when three \textbf{q}-points are summed to find a fourth one.

For the numerical approximation of the delta function, we employ an adaptive Gaussian method.
In this scheme, the delta functions appearing in the scattering processes are approximated with Gaussians
\begin{equation}
%   \delta(\Omega_s(\vec{q}) - W) \rightarrow \frac{e^{-\big(\frac{\Omega_s(\vec{q}) - W}{2\sigma^2}\big)^2}}{\sqrt{2\pi}\sigma}
    \delta(\Omega_s(\vec{q}) - W) \rightarrow \frac{e^{-(\Omega_s(\vec{q}) - W)^2 / 2\sigma^2}}{\sqrt{2\pi}\sigma}
\end{equation}
at a frequency $W$, with a width $\sigma$ estimated according to the scattering event being computed.
In the original formulation of the method~\cite{Yates2007} and the subsequent adaptations~\cite{Li2012,Li2014,Han2022}, the width is obtained by expanding $W$ linearly with respect to the $\vec{q}$ point of one of the phonons involved in the scattering.
In this work, we opted for a different, more robust approach, where each phonon frequency involved in the scattering is expanded around its respective $\vec{q}$-point.
To first order, this means that the frequency of a phonon at a point $\vec{q}$ in the neighborhood of $\vec{q}_0$ can be expressed as
\begin{equation}
    \Omega_{s}(\vec{q}) \approx \Omega_{s}(\vec{q}_0) + \sum_\alpha \frac{\partial \Omega_{s}(\vec{q})}{\partial \vec{q}^\alpha}(\vec{q}^\alpha - \vec{q}_0^\alpha)
\end{equation}
From this extrapolation, one can "blur" phonons around each $\vec{q}$-point using Gaussians with mean $\bar{\Omega}_s(\vec{q})$ and variance $\sigma_s(\vec{q})$ computed with
\begin{align}
    \bar{\Omega}_s(\vec{q}) &= E[\Omega_s(\vec{q})] = \Omega_s(\vec{q}_0) \\
    \sigma_{s}^2(\vec{q}) &= E[(\Omega_{s}(\vec{q}) - \bar{\Omega}_{s}(\vec{q}))^2] \\
    &= \bigg|\frac{\partial \Omega_{s}(\vec{q})}{\partial \vec{q}}\bigg|^2 \mid \vec{q} - \vec{q}_0 \mid^2
\end{align}
where $E[X]$ is the expectation of $X$.
The width $\sigma$ can then be obtained from the convolution of all ``blurred'' phonons involved in a specific process, giving
\begin{equation}
    \sigma = \sqrt{\sigma_{s_1}^2(\vec{q}_1) + \sigma_{s_2}^2(\vec{q}_2) + \sigma_{s_3}^2(\vec{q}_3)}
\end{equation}
for third-order processes, and
\begin{equation}
    \sigma = \sqrt{\sigma_{s_1}^2(\vec{q}_1) + \sigma_{s_2}^2(\vec{q}_2) + \sigma_{s_3}^2(\vec{q}_3) + \sigma_{s_4}^2(\vec{q}_4)}
\end{equation}
for fourth-order processes.
%Compared to approach involving group velocity differences~\cite{Han2022,Yates2007,Li2012,Li2014}, our approach does not break the symmetry of the scattering matrix.
Compared to the approach involving group velocity differences~\cite{Han2022,Yates2007,Li2012,Li2014}, our approach respect the symmetries of the scattering matrix, and allow to keep it symmetric positive definite.
%\ollecomment{Ensuring equations \ref{eq:scatmat la2} 67 hold numerically}
Moreover, the individual phonon broadening parameters $\sigma_{s}(\vec{q})$ can be precomputed at the beginning of the calculation with other harmonic properties.

\subsection{Iterative solution to the collective diagonal contribution}

The collective contribution to the thermal conductivity tensor poses a computational challenge, as it requires the diagonalization of the scattering matrix $\boldsymbol{\Xi}$. An alternative formulation of eq.(\ref{eq:kappa diag bte}) proves to be advantageous~\cite{Barbalinardo2020}
\begin{equation}
    \boldsymbol{\kappa}^{\mathrm{d}} = \frac{1}{V} \sum_\vec{q}\sum_s c_s(\vec{q}) \vec{v}_\vec{q}^s \otimes \vec{F}_s(\vec{q})
\end{equation}
where
\begin{equation}
    \vec{F}_s(\vec{q}) = \boldsymbol{\Xi}^{-1} \vec{v}_\vec{q}^s
\end{equation}
This formulation circumvents direct matrix inversion, by focusing on computing the vectors $\vec{F}_s(\vec{q})$, which can be limited to irreducible \textbf{q}-points.
Using the Neumann series for matrix inversion, $(\vec{1} - \boldsymbol{\Xi})^{-1} = \sum_{n=0}^\infty \boldsymbol{\Xi}^{n}$, and suitable reordering, an iterative method for computing $\vec{F}_s(\vec{q})$ is obtained
\begin{align}
\label{eq:iterative}
    \vec{F}_s^0(\vec{q}) =& \frac{1}{2\Gamma_s^M(\vec{q})} \vec{v}_\vec{q}^s \\
\label{eq:iterative 2}
    \vec{F}_s^{n+1}(\vec{q}) =& \vec{F}_s^0(\vec{q}) - \frac{1}{2\Gamma_s^M(\vec{q})} \sum_{\vec{q}_2 s_2} \boldsymbol{\Xi}(\vec{q}s, \vec{q}_2 s_2) \vec{F}_{s_2}^n(\vec{q}_2)
\end{align}
This iterative approach is analogous to Omini's solution to the phonon Boltzmann equation~\cite{Omini1996}.
In our implementation, convergence of the series is improved using a mixing prefactor $\alpha$ between iterations, where $0 < \alpha < 1$.
As a trade-off between memory usage and speed, the scattering matrix is retained throughout iterations, but only rows corresponding to the $N_{\mathrm{irr}}$ irreducible \textbf{q}-points are stored.
This results in a manageable storage size of $3N_\mathrm{at} N_{\mathrm{irr}} \times 3 N_{\mathrm{at}}N_q$, \emph{independent of the scattering order considered}, thereby avoiding the memory overhead associated with fourth-order terms when storing scattering processes independently~\cite{Han2022,Han2023} and allowing the use of the BLAS linear algebra library\cite{blas} to perform the matrix multiplication in eq.(\ref{eq:iterative 2}).

\subsection{Improving the computational cost}

Calculating thermal conductivity can incur significant computational costs, especially when considering fourth-order interactions.
In a na\"{\i}ve implementation, the computation of third-order interactions scales as $O(N_i\times N_q\times N_s^3)$, where $N_i$ is the number of irreducible \textbf{q}-points in the Brillouin zone, $N_q$ is the number of points in the full grid, and $N_s$ is the number of modes.
For fourth-order scattering, the scaling is even more demanding, at $O(N_i \times N_q^2 \times N_s^4)$.

In this subsection, we will demonstrate techniques to mitigate this computational cost.

\subsubsection{Computing the scattering amplitude}

The most time-consuming part of computing thermal conductivity is the calculation of the scattering matrix elements, which are needed for a large number of triplets or quartets of \textbf{q}-points and modes.
To reduce this computational cost, we divide the calculation into two steps.
First, once a triplet of \textbf{q}-point is selected, we Fourier transform the third-order IFC in reciprocal space, without projecting on mode:
\begin{equation}
\label{eq:scatmat la1}
    \widetilde{\Psi}_{ijk}^{\alpha\beta\gamma}(\vec{q}_1, \vec{q}_2, \vec{q}_3) = \sum_{\mu\nu} \frac{\Psi_{ijk}^{\alpha\beta\gamma}(\mu,\nu)}{\sqrt{M_i M_j M_k}} e^{-i (\vec{q}_2 \vec{R}_{\mu,j} + \vec{q}_3\vec{R}_{\nu,k})}
\end{equation}
Then, for each triplet of modes, corresponding to this triplet of \textbf{q}-points, the third-order IFC in reciprocal space are projected onto the phonon modes using
\begin{equation}
\label{eq:scatmat la2}
    \Psi_{\vec{q}_1\vec{q}_2\vec{q}_3}^{s_1 s_2 s_3} = \widetilde{\boldsymbol{\Psi}}(\vec{q}_1,\vec{q}_2,\vec{q}_3) \times \widetilde{\boldsymbol{\varepsilon}}_{s_1}(\vec{q}_1) \times \widetilde{\boldsymbol{\varepsilon}}_{s_2}(\vec{q}_2) \times \widetilde{\boldsymbol{\varepsilon}}_{s_3}(\vec{q}_3)
\end{equation}
with $\widetilde{\boldsymbol{\varepsilon}}_{s}(\vec{q}) = \boldsymbol{\varepsilon}_{s}(\vec{q}) / \sqrt{\Omega_{s}(\vec{q})}$.
This second step can be significantly accelerated by recognizing that it can be formulated as matrix-vector multiplications, allowing us to use optimized routines.
The same approach can be applied to the fourth-order scattering matrix elements.

\subsubsection{Irreducible triplet and quartet}

To reduce both the time and memory cost of the calculations, it is essential to exploit the symmetry properties of the scattering matrix elements. 
These elements exhibit specific symmetries under permutations of both the q-points and mode indices~\cite{Chaput2011,Togo2023}
\begin{equation}
\begin{split}
    \Psi_{P(q_1,q_2,q_3)}^{P(s_1,s_2,s_3)} = \Psi_{q_1,q_2,q_3}^{s_1,s_2,s_3} \;\; \forall P \in \mathcal{P}^{(3)} \\
    \Psi_{P(q_1,q_2,q_3,q_4)}^{P(s_1,s_2,s_3,s_4)} = \Psi_{q_1,q_2,q_3,q_4}^{s_1,s_2,s_3,s_4} \;\; \forall P \in \mathcal{P}^{(4)}
\end{split}
\end{equation}
where $\mathcal{P}^{(3)}$ represents the set of all the permutations of a triplet and $\mathcal{P}^{(4)}$ is the set of all permutations of a quartet.
Utilizing these symmetries for any irreducible triplet reduces the number of third-order elements by about half and the fourth-order elements by about a factor of six.

Furthermore, the number of elements can be further reduced by employing the symmetry operations of the crystal structure. 
For a rotation $R$ belonging to the set of crystal symmetry operations $\mathcal{R}$ expressed for reciprocal space, the invariance is expressed as~\cite{Chaput2011,Togo2023}
\begin{equation}
\begin{split}
    \Psi_{R\vec{q}_1,R\vec{q}_2,R\vec{q}_3}^{s_1,s_2,s_3} =& \Psi_{\vec{q}_1,\vec{q}_2,\vec{q}_3}^{s_1,s_2,s_3} \\
    \Psi_{R\vec{q}_1,R\vec{q}_2,R\vec{q}_3,R\vec{q}_4}^{s_1,s_2,s_3,s_4} =& \Psi_{\vec{q}_1,\vec{q}_2,\vec{q}_3,\vec{q}_4}^{s_1,s_2,s_3,s_4}
\end{split}
\end{equation}

In our implementation, once a triplet or quartet of q-points is selected, we check the possibility of reduction based on the aforementioned symmetries. 
If the triplet or quartet is reducible, the calculation of scattering matrix elements is skipped. 
The integration weights of the remaining irreducible triplets or quartets are adjusted to reflect their multiplicity accordingly.

\subsubsection{Monte-Carlo integration for the scattering rates}

Despite the improvements brought by the linear algebra formulation of the scattering matrix elements and the irreducible triplet and quartet, the computational cost remains significant due to the large number of elements involved, especially for 4th order scattering.

This cost can be greatly reduced by recognizing that the computation of $\boldsymbol{\kappa}$ can be divided into distinct integrations. 
The first (outer) integration pertains to the contribution of each q-point to the thermal conductivity, and can be written as a weighted sum over the irreducible q-points
\begin{equation}
    \boldsymbol{\kappa} = \sum_{\vec{q}^{\mathrm{irr}}} w(\vec{q}^{\mathrm{irr}}) \boldsymbol{\kappa}(\vec{q}^{\mathrm{irr}}) 
\end{equation}
where $w(\vec{q}^{\mathrm{irr}})$ is the integration weight of the irreducible q-point $\vec{q}^{\mathrm{irr}}$ and $\boldsymbol{\kappa}(\vec{q}^{\mathrm{irr}})$ represents the contribution of this q-point to the thermal conductivity.

For each irreducible point and each vibrational mode, additional inner integrations are required to compute either the lifetime or the memory kernel at the isotopic, three-phonon, and/or four-phonon levels. 
Typically, a full grid is used for these integrations. 
However, the q-point grid densities needed to converge the different integrations are not necessarily the same. 
Specifically, the grid densities required to converge the linewidths are usually much lower than those needed for the outer integration for the thermal conductivity, as we will demonstrate in the applications section. 
Consequently, we implement a scheme to decouple these integrations, using a Monte-Carlo method on the grids for the inner integrals. 
This decoupling significantly reduces the computational cost, with the only drawback being the introduction of (controllable, numerical) noise into the results.

\begin{figure}
    \centering
    \includegraphics[width=\columnwidth]{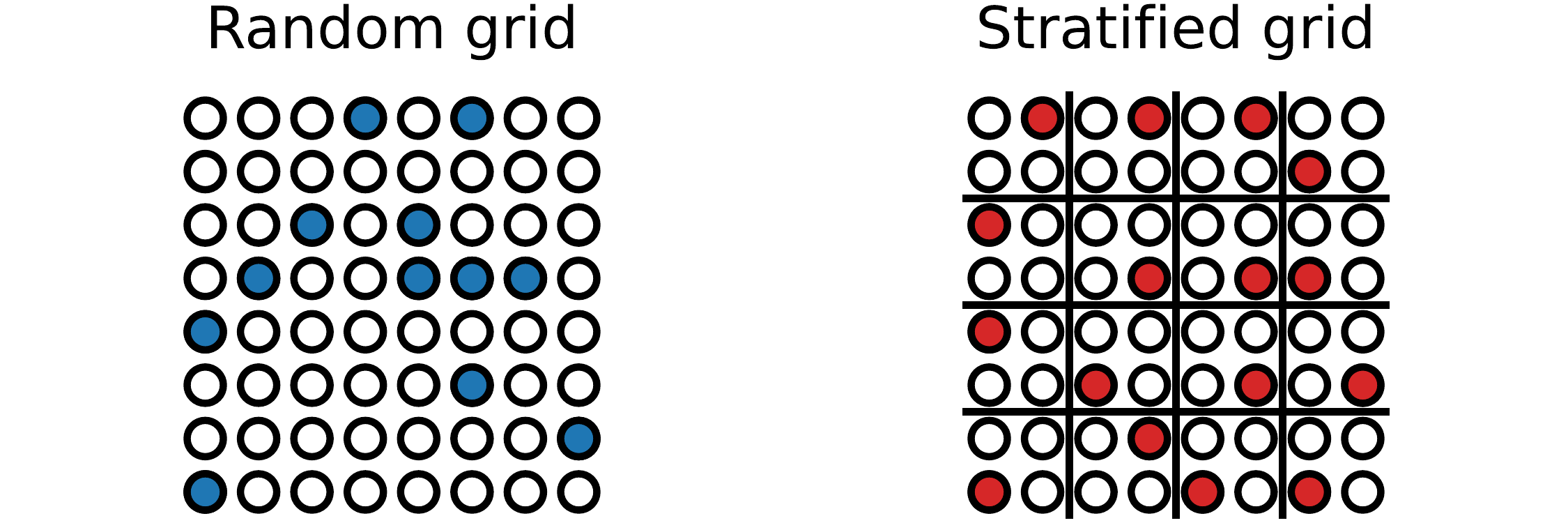}
    \caption{Sketch of the method used to select q-points for the Monte-Carlo integrations. 
    In this example, the full grid is $8\times8$ and we use a $4\times4$ grid for the Monte-Carlo.
    For all grids, empty circles represent a point on the $8\times8$ full grid, while the colored circles represent the point selected for the Monte-Carlo integration.
    When the Monte-Carlo grid points are selected randomly, the distribution of points is less uniform and some part of reciprocal space can be left empty while others are ``bunched'' and oversampled.
    Moreover, the same point can be selected several times, reducing even more the uniformity of the distribution.
    With our stratified approach, represented on the bottom right, the subdivisions of the full grid ensure that reciprocal space is more uniformly sampled.}
    \label{fig:MonteCarlo}
\end{figure}

Initially, a dense grid is generated, and all necessary harmonic quantities are computed on it. 
This grid, which will be used for the thermal conductivity integration, is referred to as the full grid. 
Then, for each scattering integration, we compute the contributions from a randomly selected subset of points of the dense grid, termed the Monte-Carlo grid. 
To enhance the convergence of the integrals with respect to the Monte-Carlo grid densities, the points are not selected entirely at random but rather using a stratified approach. 
In this approach, the full grid is subdivided into smaller sections, and points are randomly selected within these subdivisions. 
This ensures that the Monte-Carlo grid samples the reciprocal space more uniformly, as shown in Fig. \ref{fig:MonteCarlo}, thereby reducing the variance of the results.

It should be noted that a similar method has been proposed based on a maximum likelihood justification~\cite{Guo2024}.
However, while the maximum likelihood approach is based on the RTA, our Monte-Carlo integration method is agnostic to the quantity computed and can be used for any approximation derived in this work. 
Moreover, our stratification step ensures that if the Monte-Carlo and full grids have the same densities, all points on the full grid are used in the Monte-Carlo grid, making the inner integrations deterministic and yielding results equivalent to those obtained using the full grid in all steps.

Empirically we have found that this approach is more delicate for the computation of the off-diagonal terms of the scattering matrix.
Indeed, if the terms that couple \textbf{q}-points $(\vec{q}_1, \vec{q}_2)$ are skipped by the Monte-Carlo scheme, then the corresponding entries of the matrix will be empty, thus neglecting coupling between the corresponding modes.
While this should not be a problem for most systems, where collective effects contribute only a small fraction of the thermal conductivity, this neglect can be dramatic for materials such as graphene, where the collective contribution is the dominant source in heat transport~\cite{Fugallo2014}.
Fortunately, the scattering matrix should respect some symmetries that can be enforced to alleviate this problem.
For instance, given a rotation $\vec{R}$ belonging to the little group of $\vec{q}_1$, the scattering matrix should respect the relation
\begin{equation}
    \Xi_{s_1 s_2}(\vec{R} \vec{q}_1, \vec{R} \vec{q}_2) = \Xi_{s_1 s_2}(\vec{q}_1, \vec{q}_2)
\end{equation}
In our implementation, we impose this symmetry in a way that fills the neglected entries of the scattering matrix with the average of the values of other equivalent entries.

\section{Applications}
\label{sec:applications}

In this section, we demonstrate the performance and precision of the formalism and implementation detailed in this paper through various applications.
To provide a comprehensive overview, we selected systems that represent the diverse regimes of thermal conductivity of thermal conductivity.

\begin{figure}
    \centering
    \includegraphics[width=\columnwidth]{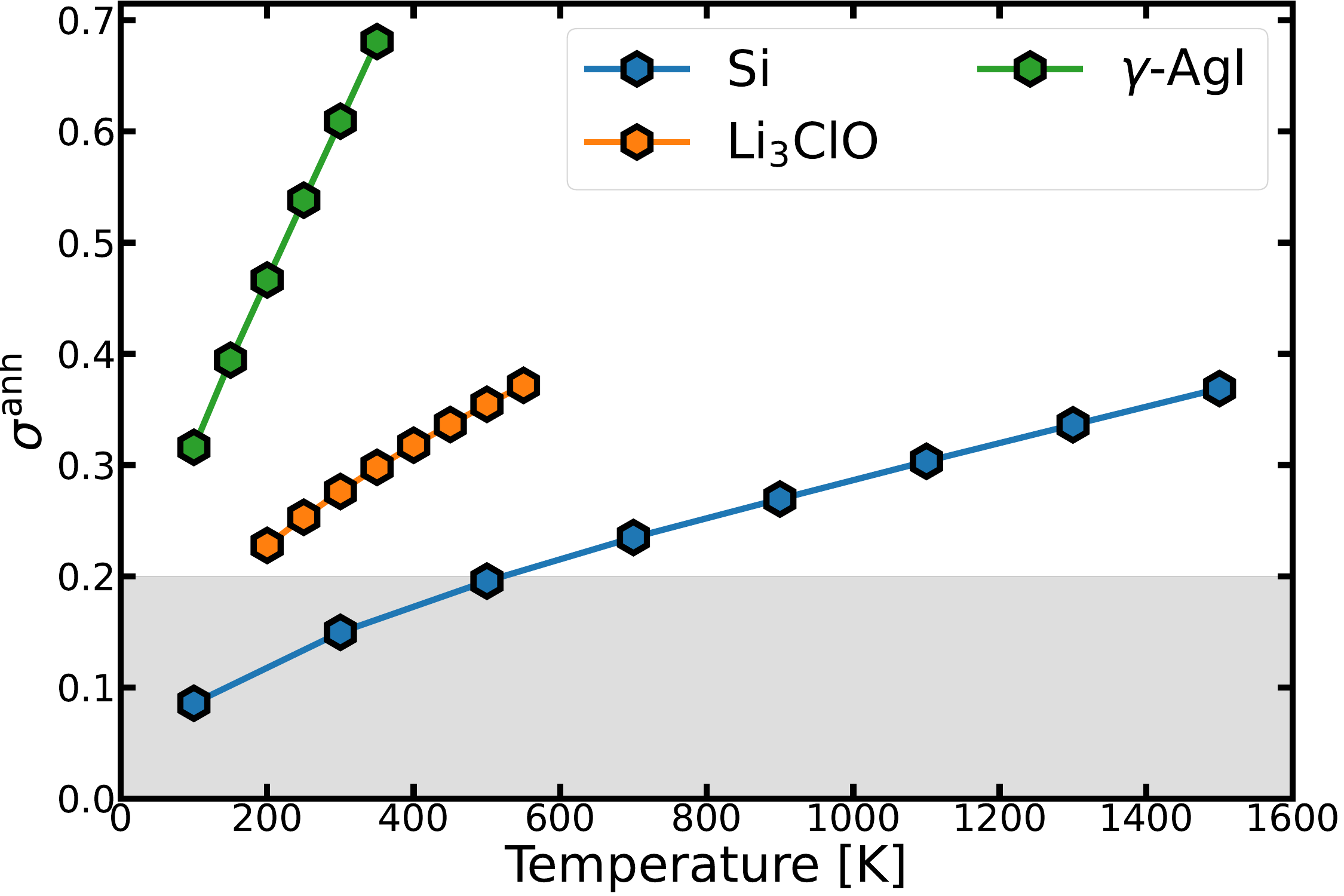}
    \caption{Anharmonicity measure of the materials studied in this work. The gray zone indicates low anharmonicity.}
    \label{fig:anharmonicity}
\end{figure}

To quantify the different regimes of anharmonicity spanned by our sample materials, we use the anharmonicity measure introduced by Knoop \textit{et al}~\cite{Knoop2020}
\begin{equation}
    \sigma^{\mathrm{anh}}(T) = \sqrt{\frac{\sum_{i\alpha} \braket{(\delta f_i^\alpha)^2}}{\sum_{i\alpha} \braket{(f_i^\alpha)^2}}}
\end{equation}
which, in the context of the mode-coupling theory, becomes a measure of the dissipative component of the dynamics of the system~\cite{Castellano2023}.
In Fig.~\ref{fig:anharmonicity}, we plot the anharmonicity measure for the different materials considered in this work.
Going from low to high anharmonicity, our example systems are silicon ($\sigma^{\mathrm{anh}}(T=300K)=0.15$), Li\textsubscript{3}ClO ($\sigma^{\mathrm{anh}}(T=300K)=0.28$) and $\gamma$-AgI ($\sigma^{\mathrm{anh}}(T=300K)=0.61$).

\subsection{Framework to compute the thermal conductivity and numerical details}
\label{sec:framework}

The key components for computing the thermal conductivity tensor are the generalized IFC, which must be derived from MD, or PIMD for systems with significant nuclear quantum effects.
Performing MD simulations for each temperature with an \textit{ab-initio} description of the Born-Oppenheimer surface incurs substantial computational expense, even when using DFT.
To mitigate this cost, we propose a comprehensive framework utilizing machine-learning interatomic potentials (MLIP) as surrogates for the \textit{ab-initio} Born-Oppenheimer surface in MD simulations.

%   \begin{figure}
%       \centering
%       \includegraphics[width=\columnwidth]{FlowchartTc.pdf}
%       \caption{Obviously work in progress, just to have an idea of the size and things like this}
%       \label{fig:flowchart_tc}
%   \end{figure}

% Our framework consists of three main steps, as illustrated in Fig.\ref{fig:flowchart_tc}.
Our framework consists of three main steps.
First, starting from the crystal structure, a MLIP is trained using a self-consistent approach.
In this method, the MLIP is iteratively trained and used to generate configurations, which are then added to a dataset.
It should be noted that configurations are added randomly, without any accuracy criterion, in order to sample uniformly the canonical ensemble of the systems.
Using a variational principle, it can be shown that this approach yields an optimal MLIP according to the Kullback-Leibler divergence~\cite{Castellano2022}, enhancing accuracy for equilibrium properties at the expense of extrapolation capacity.

Once the MLIP is prepared, MD simulations in the NPT ensemble are conducted for each desired temperature to determine the system equilibrium volumes.
This step is crucial because thermal expansion significantly affects the renormalization of phonon frequencies, thus impacting thermal conductivity.
The equilibrated cell is then used for MD simulations in the NVT ensemble and configurations from these simulations are extracted to compute the generalized IFCs, which are subsequently used to calculate the thermal conductivity.

It should be noted that while we use classical MD in the remaining of this work, this scheme can easily be adapted to systems where nuclear quantum effects are important, by simply replacing classical MD with path-integral MD.

\subsection{Computational parameters}

For all applications, DFT calculations are performed with the Abinit suite~\cite{Gonze2020,Romero2020}.
The MLIP employed in this work uses the Moment Tensor Potential framework~\cite{Shapeev2016,Novikov2021}, with a level 22 and a $6$~\r{A} cutoff for every material.
MD simulations are executed with the LAMMPS software~\cite{Thompson2022}, utilizing the GJF integrator~\cite{GrnbechJensen2013} for Langevin dynamics.
Finally, the computation of the generalized IFCs and the thermal conductivity tensor is carried out using the TDEP package~\cite{Knoop2024}.
More information on the computational details can be found in Appendix \ref{app:param}.

For the NPT and NVT molecular dynamics, we used a $4\times4\times4$ supercell of Li\textsubscript{3}ClO and AgI, totaling 350 and 512 atoms and a $3\times3\times3$ supercell for Si, with 216 atoms.
The cutoff for the second order generalized IFC was set at half the size of the supercell for all systems.
For the third order, we used a cutoff of $7.3$, $6.4$ and $7.0$~\r{A} for Si, Li\textsubscript{3}ClO and AgI, respectively, while we used $4.0$, $3.0$ and $5.0$~\r{A} at the fourth order.
These parameters were selected after careful convergence of the thermal conductivity tensor to below 1\%.

All calculations for the thermal conductivity tensor were performed on the Lucia supercomputer of the CECI consortium in the Walloon region of Belgium.
Each node in this cluster is equipped with two AMD EPYC 7763 processors, each featuring 64 cores with a clock speed of 2.45 GHz.
For all timing measurements presented in this study, computations were conducted using a single node.

A direct comparison with experimental measurement could suffer from inaccuracies due to underlying DFT or MLIP biases.
To better assess the accuracy of our theory for all systems, we also computed the thermal conductivity using approach-to-equilibrium molecular dynamics (AEMD)~\cite{Sheng2022,Lampin2013,Martin2022,Lambrecht2024,Melis2014}.
Being a non-equilibrium MD method, AEMD includes all orders of anharmonicity in its description of heat transport, with the drawback of a strong size dependence and absence of nuclear quantum effects.
More details on our AEMD simulations are provided in appendix \ref{app:aemd}.

\subsection{Silicon}

We start our applications with silicon, a critical material in the semiconductor industry.
Its thermal conductivity has been extensively researched both theoretically and experimentally, making it an ideal candidate for benchmarking the approaches developed in this work.
Silicon is typically considered to exhibit low anharmonicity, and perturbation theory has been shown to accurately reproduce its transport properties, at least below room temperature.
Due to the large mean-free path observed in this system at low temperature, which would require very large simulation boxes to obtain convergence, we only applied AEMD at $700$, $1100$ and $1500$~K.

\begin{figure}
    \centering
    \includegraphics[width=\columnwidth]{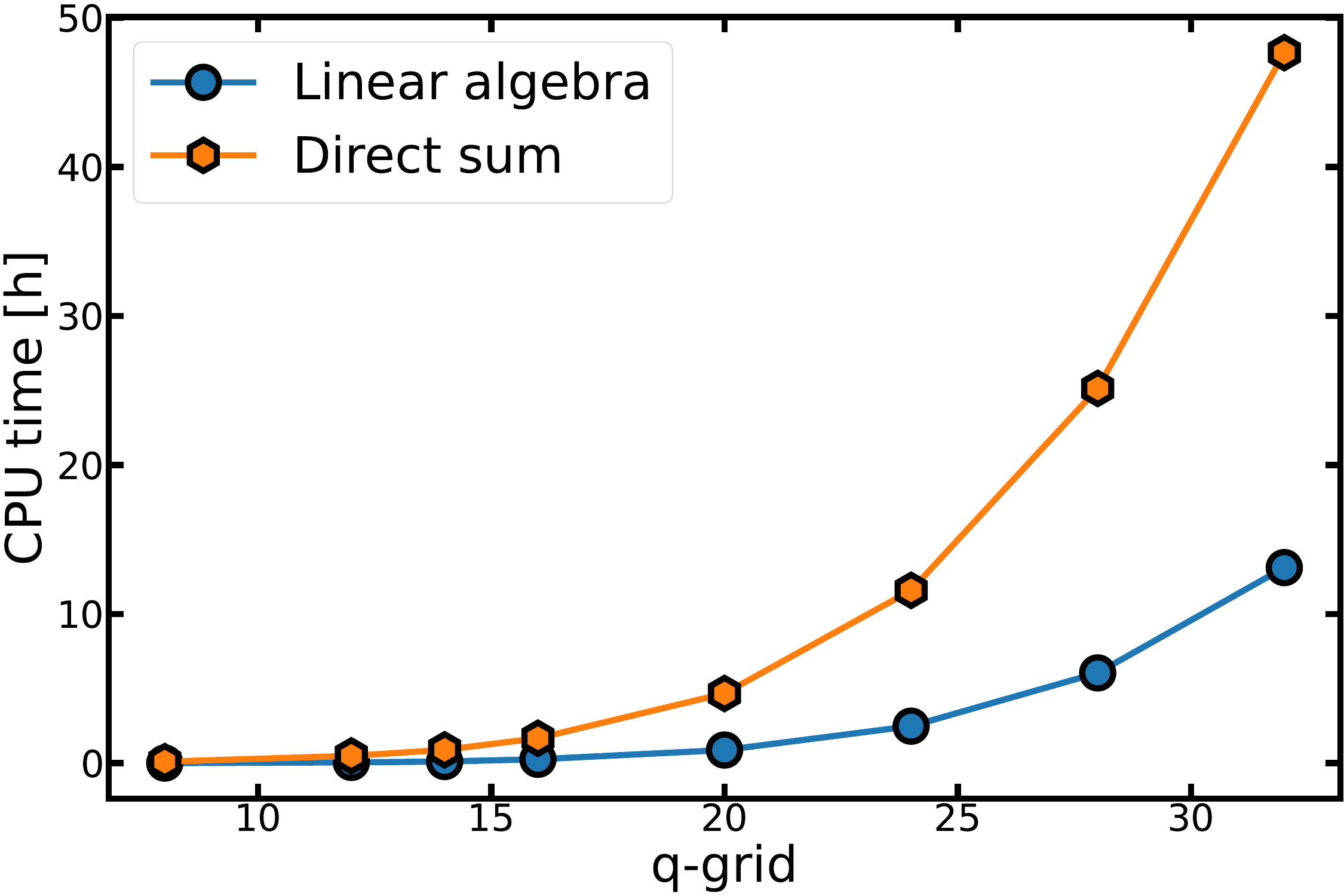}
    \caption{Computational cost for the thermal conductivity of silicon with respect to the full grid density using third order scattering.
    The blue dots show timings computed using the linear algebra scheme of eq.(\ref{eq:scatmat la1}) and (\ref{eq:scatmat la2}) while the timings using standard formulation of eq.(\ref{eq:scatmat old}) are denoted by the orange hexagon.}
    \label{fig:time_newvsold}
\end{figure}

We begin with a demonstration of the speed-up provided by using the linear algebra formulation of the scattering matrix of eq.(\ref{eq:scatmat la1}) and (\ref{eq:scatmat la2}).
Figure \ref{fig:time_newvsold} shows that the new formulation allows for a drastic reduction of the CPU time, dividing for instance by more than 3 the computational cost with a \textbf{q}-point grid of $35\times35\times35$.
It should be noted that, as the number of atoms in the unitcell or the order of the scattering matrix element increases, so does the speed-up.

\begin{figure}
    \centering
    \includegraphics[width=\columnwidth]{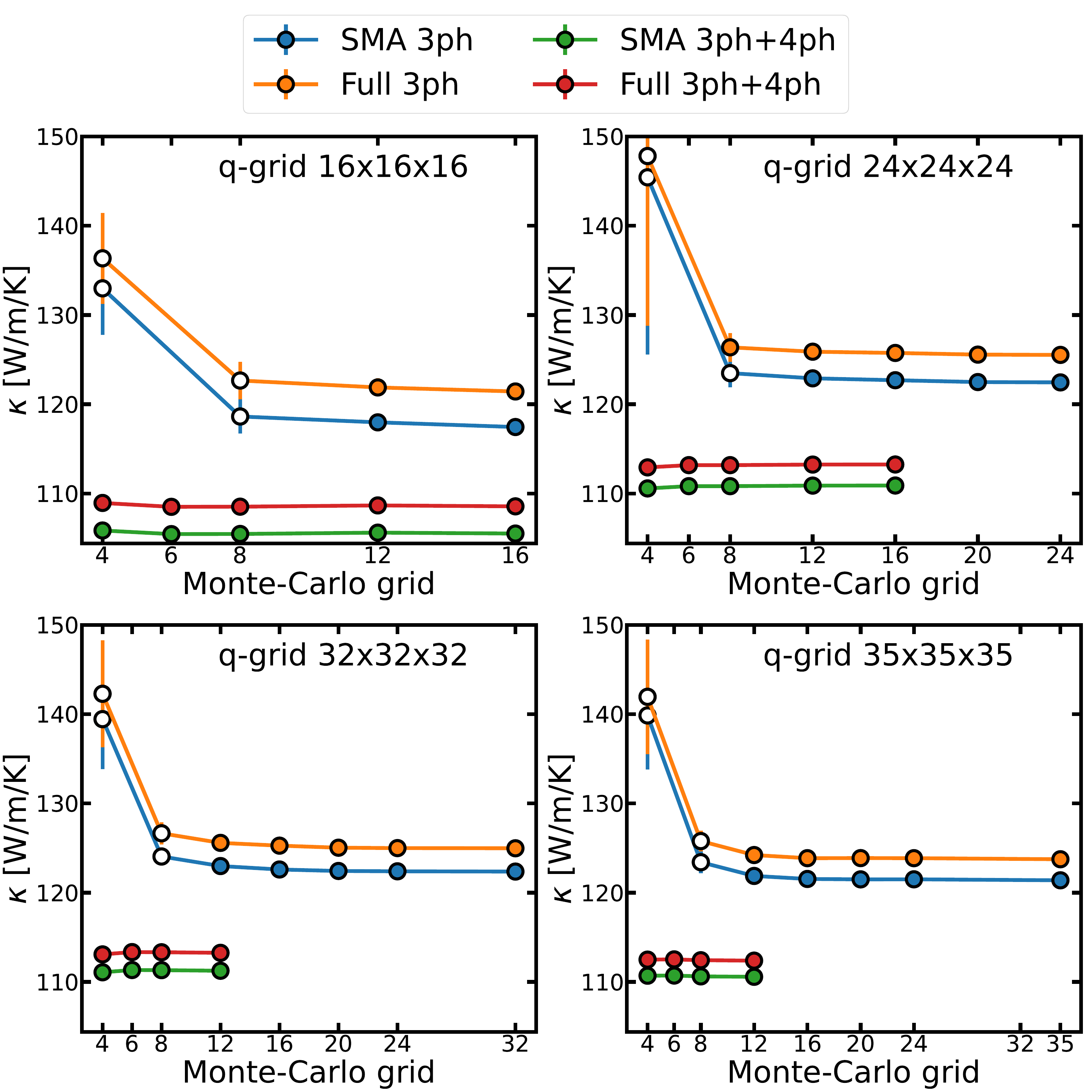}
    \caption{Convergence of the thermal conductivity with respect to the Monte-Carlo grid density for several full grid densities.
    Each point is the average over 10 calculations, except for the $12\times12\times12$ and $16\times16\times16$ grid with fourth-order, and the error bars indicate the standard deviation.
    Results with fourth-order including uses the full grid as a Monte-Carlo grid for the third order.
    SMA denotes results using the single mode approximation and Full denotes results computed using the full scattering matrix. Empty symbols indicates points with an error greater than 1~W/m/K compared to the largest grid.}
    \label{fig:conv_si_rta}
\end{figure}

We continue with the improvement brought by the Monte-Carlo integration scheme.
Figure \ref{fig:conv_si_rta} illustrates the convergence of the thermal conductivity with respect to the Monte-Carlo grid density, across several full grids.
These results clearly demonstrate the decoupling between thermal conductivity and scattering integrations.
For all the grids considered, an $8\times8\times8$ Monte-Carlo grid achieves an error of less than 1\% and a standard deviation of less than 1 W/m/K compared to scattering integration on the full grid.
Notably, this convergence is independent of the type of approximation used, validating the effectiveness of our Monte-Carlo scheme even for calculations beyond the single mode approximation.

\begin{figure}
    \centering
    \includegraphics[width=\columnwidth]{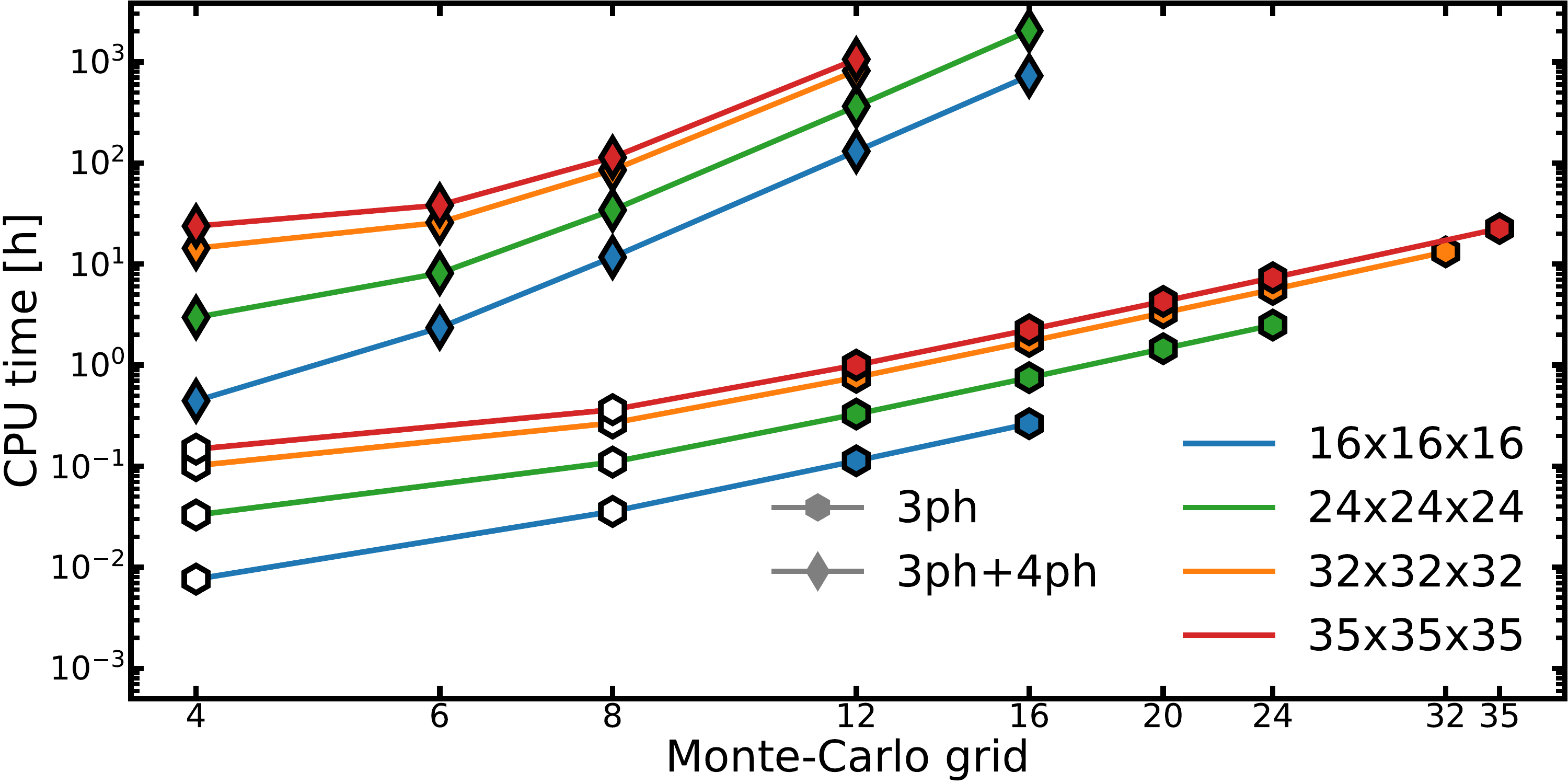}
    \caption{Computational cost for the thermal conductivity of silicon with respect to the Monte-Carlo grid density for different full grid densities.
    For the calculation involving fourth-order scattering, the Monte-Carlo grid for the third-order contribution is set to the same density as the full grid.
    Each point is the average over 10 calculations (variance smaller than symbols), except for the $12\times12\times12$ and $16\times16\times16$ grid with fourth-order. Empty symbols indicates points with an error greater than 1~W/m/K compared to the largest grid.}
    \label{fig:time_si_rta}
\end{figure}

The efficiency of this scheme is further highlighted in Fig.\ref{fig:time_si_rta}, where it is shown that using a Monte-Carlo grid of $12\times12\times12$ \textbf{q}-points can reduce the wall time by an order of magnitude compared to a full grid.
This acceleration becomes even more pronounced when fourth-order scattering processes are included.
In such cases, even a $4\times4\times4$ grid is sufficient to converge the 4th order contribution to thermal conductivity to less than 1 W/m/K, cutting computational cost by several orders of magnitude compared to full grid calculations.

The rapid convergence with smaller grids at the fourth-order can be attributed to the combinatorially large number of scattering process it involves, combined with stochastic error cancellation.
For instance, in a system like silicon, a $4\times4\times4$ grid incorporates a number of interactions on the same order of magnitude as the number of three-phonon processes within a $35\times35\times35$ \textbf{q}-point grid.
While the specific grid densities required to converge the thermal conductivity tensor vary by system, this example demonstrates the significant computational acceleration achievable with our decoupling scheme.

%\begin{figure}
%    \centering
%    \includegraphics[width=\columnwidth]{UncertaintySi.pdf}
%    \caption{Evolution of the error in the thermal conductivity (top) and variance of the results (bottom) with respect to the Monte-Carlo grid density.
%    Each point is the average the average over 10 calculations.
%    Computation are done using only third order scattering.}
%    \label{fig:unc_si_rta}
%\end{figure}

\begin{figure}
    \centering
    \includegraphics[width=\columnwidth]{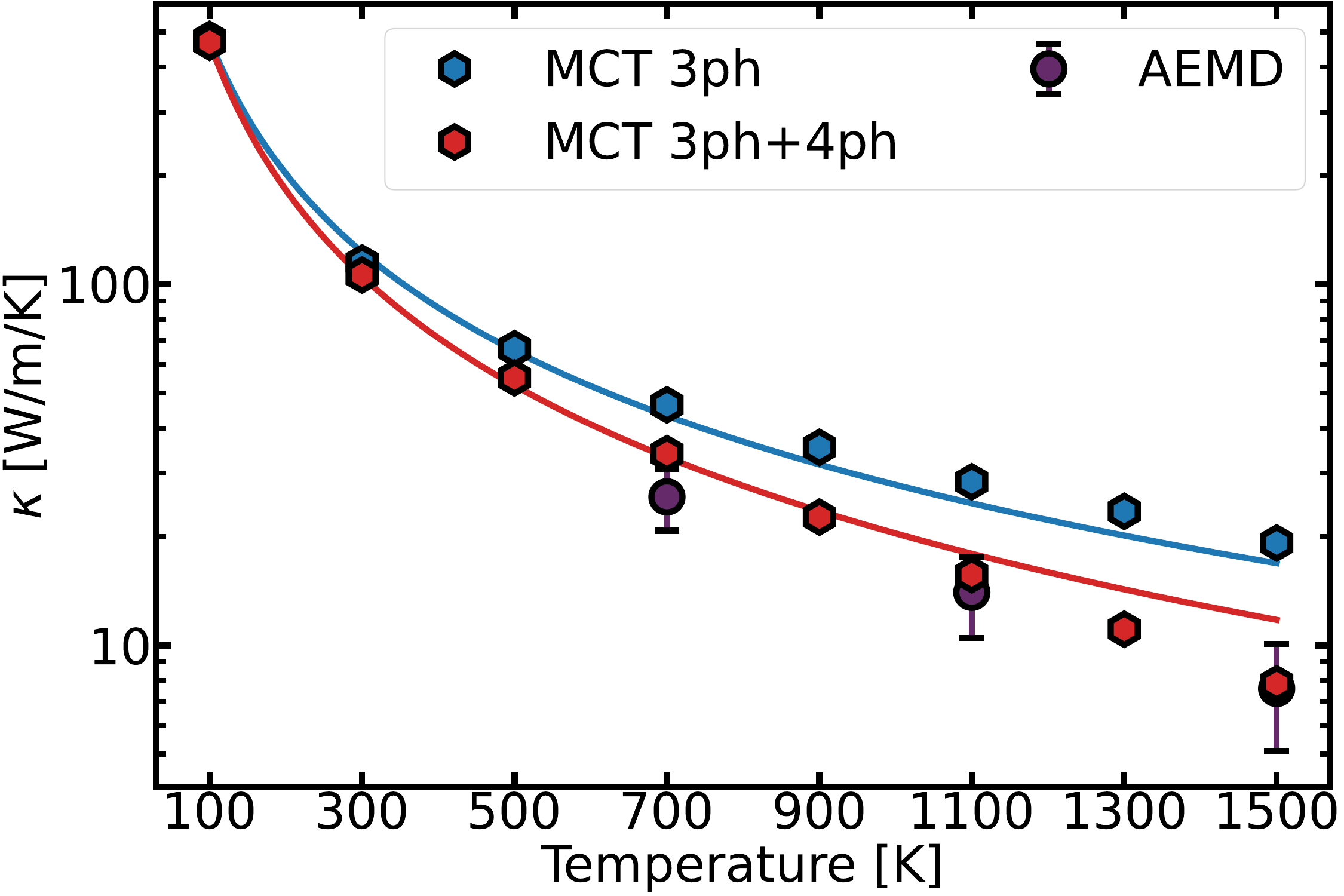}
    \caption{Evolution of the thermal conductivity of silicon between 100 and 1500~K.
    Lines are fit of the results according to the model $\kappa(T) = A / T^C$, with $A$ and $C$ as fitting parameters.}
    \label{fig:tc_si}
\end{figure}

Using a $35\times35\times35$ full grid along with a $16\times16\times16$ and a $8\times8\times8$ Monte-Carlo grids for third- and fourth-order scattering, we computed the temperature dependence of silicon's lattice thermal conductivity.
The results are displayed in Fig.~\ref{fig:tc_si}.
Our findings demonstrate good agreement with both experimental~\cite{Fulkerson1968} and theoretical results from the literature~\cite{Han2022,Gu2020}.
Notably, we observe an increasing significance of fourth-order scattering with rising temperatures, a trend corroborated by recent studies~\cite{Gu2020} and also reproduced with our AEMD simulations, with which the mode-coupling theory agrees very well.

It can be noted that our results show some discrepancy with previous results from the literature, both experimental~\cite{Inyushkin2004} and theoretical~\cite{Feng2016,Broido2007}.
Such discrepancies can be attributed to the MLIP used in this work.
It has been shown that MLIP errors on the prediction of forces can introduce an underestimation of the thermal conductivity tensor through the introduction of an effective scattering channel~\cite{Wu2024,Zhou2025}.
While perturbative theory can be quite insensitive to the forces error, the mode-coupling theory is build on averages of thermally excited configurations and a similar mechanism could be at play in this case.
However, setting up a correction scheme, such as the ones introduced for (non-)equilibrium molecular dynamics methods, is out of the scope of this work, and the agreement between AEMD and the mode-coupling theory provides a clear validation of the latter.

\subsection{Li\textsubscript{3}ClO}

The second system we studied is Li\textsubscript{3}ClO, an anti-perovskite with a rich Lithium composition that makes it a candidate future generation electrolyte in solid-state batteries~\cite{Braga2014,L2014}.
The thermal conductivity of this system has been studied using both lattice dynamics approaches, in the perturbative regime~\cite{Fiorentino2023,Pegolo2022}, and molecular dynamics within the Green-Kubo formalism~\cite{Pegolo2022}.
Li\textsubscript{3}ClO can be considered as a system with medium anharmonicity, as can be attested by its $\sigma^{\mathrm{anh}}(T)$ going from 0.23 to 0.37 when increasing the temperature from 200 to 550~K.

For this system, we use a $24\times24\times24$ full grid alongside an $8\times8\times8$ Monte-Carlo grid for third-order scattering and a $3\times3\times3$ grid for fourth order interactions.
This setup provides converged results with a computational cost of approximately $1.5$ CPU-hours when only third-order scattering is included, and around $120$ CPU-hours when fourth-order scattering is also accounted for.
To examine the influence of generalized IFC, thermal conductivity was also calculated across all temperatures using IFCs extracted at $10$~K.
Although not exactly equivalent to purely harmonic or perturbative results, these results should give an indication of possible failures of perturbative theory.

\begin{figure}
    \centering
    \includegraphics[width=\columnwidth]{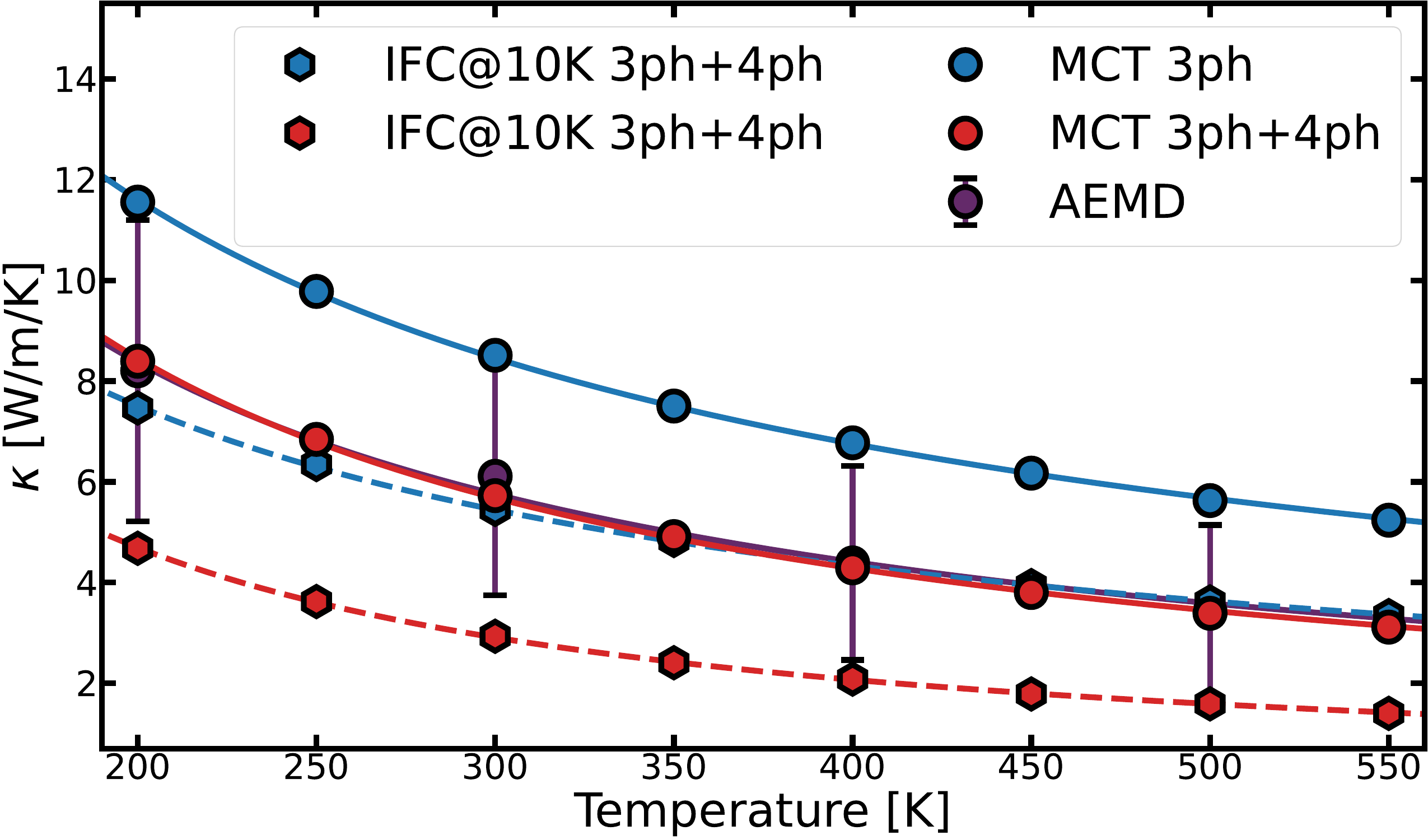}
    \caption{Evolution of the thermal conductivity of Li\textsubscript{3}ClO with respect to temperature.
    Lines are fit of the results according to the model $\kappa(T) = A / T^C$, with $A$ and $C$ as fitting parameters.
    Full lines correspond to fit of MCT and AEMD results and the dashed lines denote the fit of results with temperature independent IFC.}
    \label{fig:tc_li3clo}
\end{figure}

Our findings, summarized in Fig. \ref{fig:tc_li3clo}, reveal a significant contribution from fourth-order scattering, even at the lowest temperatures considered.
At $200$~K, the inclusion of fourth-order effects reduces thermal conductivity by approximately $37$~\%.
This impact of fourth-order scattering persists whether or not the temperature dependence of the IFCs is included in the heat transport calculations.
However, incorporating this temperature dependence is critical for an accurate description of the system's thermal conductivity.
When neglected, the third-order results yield a misleading agreement between AEMD and lattice dynamics calculations, and the inclusion of fourth-order scattering in the low T harmonic model leads to a significant underestimation of $\kappa$ across all temperatures.
In contrast, mode-coupling theory restores agreement with molecular dynamics simulations, provided that fourth-order scattering is also taken into account.

\subsection{$\gamma$-AgI}

As an example of a strongly anharmonic material, we apply our formalism to silver iodide in its zincblende $\gamma$ phase - a silver halide with applications in photovoltaic devices and solid-state batteries.
Despite its simple crystal structure, AgI, is known for its ultra-low thermal conductivity, measured to be approximately $0.4$~W/m/K at room temperature~\cite{Wang2023}.
This value is significantly overestimated by conventional perturbation theory accounting only for three phonons interactions, which predicts a thermal conductivity of $2.1$~W/m/K~\cite{Togo2015}.
Recent studies~\cite{Wang2023,Ouyang2023} have reconciled experimental and theoretical discrepancies by highlighting the critical role of phonon renormalization and high-order scattering processes.
Reference~\cite{Ouyang2023} suggested that scattering processes beyond fourth-order are necessary to accurately reproduce the ultra-low thermal conductivity of AgI, even when using anharmonic phonon theories.
However, it is important to note that their work did not rely on the mode-coupling definition of anharmonic phonons.
While they used a similar approach to MCT for calculating non-interacting phonons, the higher-order generalized IFC were computed using a different method, with all orders greater than second fitted simultaneously to molecular dynamics data instead of successively.

\begin{figure}
    \centering
    \includegraphics[width=\columnwidth]{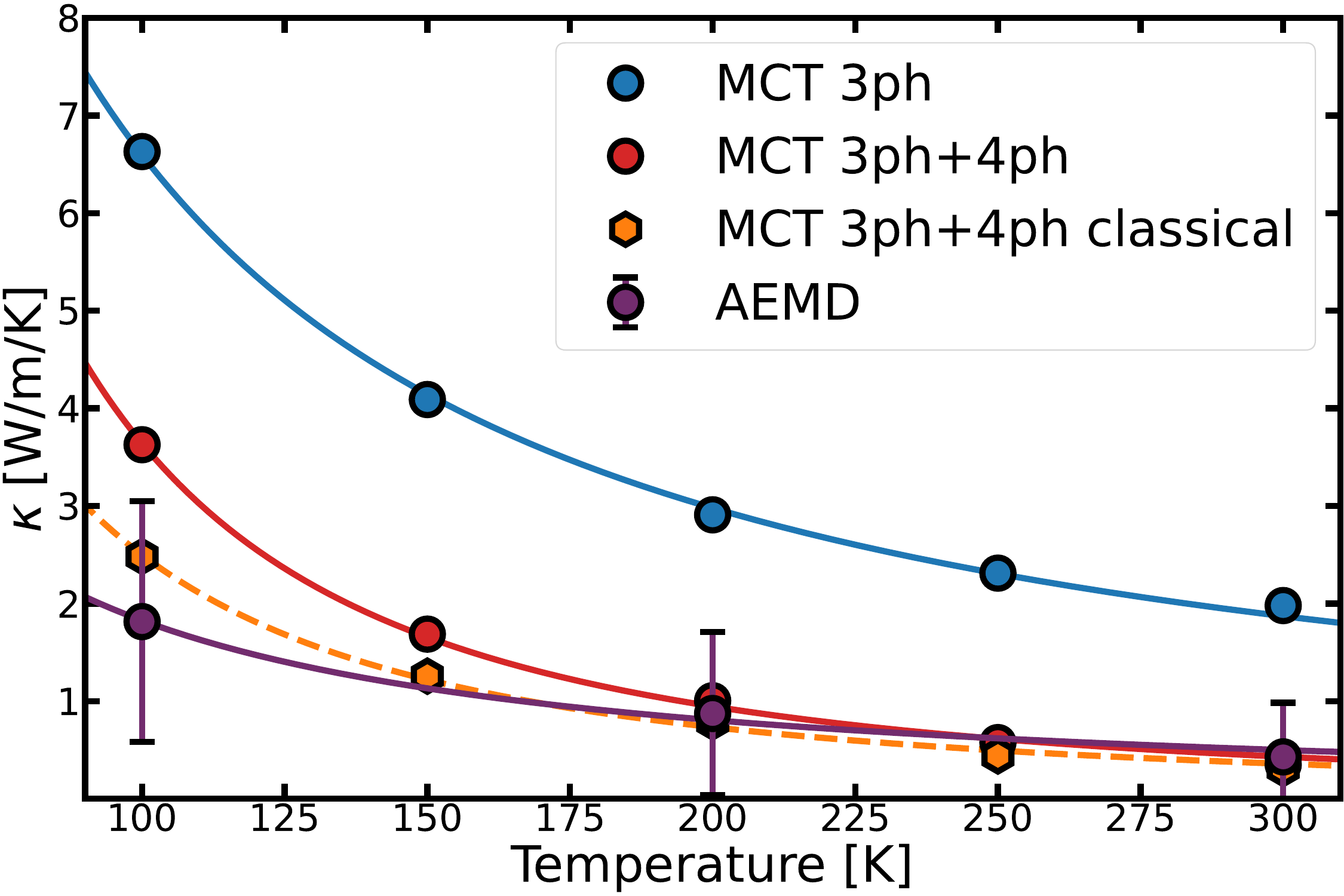}
    \caption{Evolution of the thermal conductivity of AgI with respect to temperature.
    Lines are fit of the results according to the model $\kappa(T) = A / T^C$, with $A$ and $C$ as fitting parameters.
    The dashed lines are fit of the results using classical occupation for the computation of $\kappa$.}
    \label{fig:tc_AgI}
\end{figure}

For this system, we employed a $32\times32\times32$ full grid, along with $16\times16\times16$ and $6\times6\times6$ Monte-Carlo grids for third- and fourth-order scattering, yielding a computational cost of approximately $0.5$ CPU-hours and $14$ CPU-hours, respectively.
Our results, shown in fig. \ref{fig:tc_AgI}, demonstrate the the mode-coupling theory with fourth-order scattering is sufficient to produce converged results, yielding a thermal conductivity of about $0.3$~W/m/K at room temperature, consistent with both experimental data and AEMD results.
These findings underscore the importance of a correct definition for generalized IFCs.
While second-order IFCs are crucial for accuracy, a consistent definition of higher-order IFCs is equally essential to capture the full anharmonic behavior of the system.
In the end, using a classical distribution instead of the Bose-Einstein, our mode-coupling results are able to reproduce AEMD for the whole range of temperature considered, despite the strong anharmonicity of this system. A strong advantage of our mode coupling over MD techniques is the absence of finite size artefacts and the insight in mode contributions and scattering mechanisms.

\section{Conclusion}
\label{sec:conclusion}

%The search for materials with either high or low thermal conductivity is an intense topic of research.
In recent years, the limitations of the harmonic approximation have become more and more apparent, producing a surge in the use of temperature dependent phonon theories.
In this work, we provided a detailed derivation of the theory of thermal conductivity in the framework of the mode-coupling theory of anharmonic lattice dynamics, thus justifying the use of temperature dependent phonons as a means to study heat transport in materials.

Following a summary of the mode-coupling theory for anharmonic crystals, the starting point of our derivation consists in the introduction of a consistent formulation of the heat current.
Using the Green-Kubo formula, we obtain an equation for the thermal conductivity tensor involving correlation functions for the phonon operators.
Our final result for the $\boldsymbol{\kappa}$ tensor includes single mode, collective and off-diagonal contributions, which makes it valid for a large range of systems, from simple to complex crystals with low or high anharmonicity.

Due to the extreme computational cost incurred in the computation of scattering processes, we also present numerical strategies to increase the efficiency of thermal conductivity calculations, and fully converge even the most advanced calculations with dense grids.
This acceleration is enabled by reducing the cost of each scattering processes, through the use of a linear algebra formulation of the scattering matrix elements, and by reducing the number of processes to be computed.
For this second point, we implemented both symmetry reduction and Monte-Carlo integration schemes.
All these improvements result in a drastic reduction of the computational cost, and even enable the computation of thermal conductivity with up to fourth order scattering, in complex systems for which it would otherwise be too expensive.

Finally, we apply our implementation to systems going from low to high anharmonicity.
This demonstrates the validity of the mode-coupling theory of anharmonic lattice dynamics to compute transport properties, as well as the efficiency of our implementation.

Regarding the limitations of our work, we stress the two main approximations made.
The first concerns the use of the Markovian limit, in which the full frequency-dependent phonon spectral functions are replaced by Lorentzians characterized only by their center (the frequency $\Omega_s(\vec{q})$) and width (the linewidths $\Gamma_s^M(\vec{q})$.
While this approximation is ubiquitous in the literature, the more complicated spectral functions observed in some anharmonic materials raise questions on the validity of the Markovian approximation in such cases.
One should notice that in this work, we already derived eq.(\ref{eq:off diagonal nonmarkov}), which goes beyond the Markovian limit.
The other main approximation is the neglect of a part of the heat current operator involving the random force and memory kernel.
As the anharmonicity of a material increases, their importance is expected to increase.
However, these contributions have never been considered, and their importance in realistic materials is unknown:
our work provides a starting point to quantify these terms.

\begin{acknowledgments}
The authors gratefully thank Florian Knoop for reviewing the implementation and for suggestions to improve the manuscript.

The authors acknowledge the Fonds de la Recherche Scientifique (FRS-FNRS Belgium) and Fonds Wetenschappelijk Onderzoek (FWO Belgium) for EOS project CONNECT (G.A. 40007563), and 
F\'ed\'eration Wallonie Bruxelles and ULiege for funding ARC project DREAMS (G.A. 21/25-11).
MJV acknowledges funding by the Dutch Gravitation program
“Materials for the Quantum Age” (QuMat, reg number 024.005.006), financed by the Dutch Ministry of Education, Culture and Science (OCW).

% CPU
Simulation time was awarded by 
by PRACE on Discoverer at SofiaTech in Bulgaria (optospin project id. 2020225411), 
EuroHPC-JU award EHPC-EXT-2023E02-050 on MareNostrum 5 at Barcelona Supercomputing Center (BSC), Spain
by the CECI (FRS-FNRS Belgium Grant No. 2.5020.11),
and by the Lucia Tier-1 of the F\'ed\'eration Wallonie-Bruxelles (Walloon Region grant agreement No. 1117545).
\end{acknowledgments}

\appendix

\section{Relation between some correlation functions}
\label{app:correlation functions}

If the definition of a classical correlation function is unambiguous, there exists an infinite number of way to define a quantum correlation function.
In this appendix, we give the relation between the Kubo correlation function used throughout this work and other types of correlation functions used to obtain some of the relations appearing in the main text.
Most of the equalities given here use properties of the Bose-Einstein distribution, in particular $n(-\omega) = -(n(\omega) + 1)$ or $e^{-\hbar\omega/\kBT} = n(\omega) / (n(\omega) + 1)$.

The first important relation is that between the Kubo correlation function and the generalized susceptibility, which stems from the fluctuation-dissipation theorem~\cite{Castellano2023}
\begin{equation}
    \chi''(\omega) = \frac{\omega}{2\kBT} G(\omega)
\end{equation}
Other important quantum correlation function are the lesser and greater ones, which is defined for two operators $A$ and $B$ as
\begin{align}
    G^<(t) =& \braket{A B(t)} \theta(t) \\
    G^>(t) =& \braket{A B(t)} \theta(-t)
\end{align}
where $\theta(t)$ is the Heaviside function.
Decomposing the correlation functions on the eigenstates of the Hamiltonian into a Lehmann representation and using the properties of the Bose-Einstein distribution, one can show that the lesser, greater and Kubo correlation functions are related by
\begin{align}
    G^<(\omega) =& \frac{\hbar\omega}{\kBT} n(\omega) G(\omega) = n(\omega) \chi''(\omega) \\
    G^<(\omega) =& \frac{\hbar\omega}{\kBT} (n(\omega) + 1) G(\omega) = (n(\omega) + 1) \chi''(\omega)
\end{align}

\section{Decoupling Kubo correlation functions}
\label{app:decoupling}

In this appendix, we formally derive the decoupling of four-point Kubo correlation functions of the form $(A B, C(t), D(t))$, where $A$, $B$, $C$ and $D$ are arbitrary operators.
In our previous work, we used the decoupling
\begin{equation}
    (A B, C(t) D(t)) \approx \big(A, C(t)\big)\big(B, D(t)\big) + \mathrm{perm}
\end{equation}
with perm being the permutations of the operators in the four-point correlation function.
However, this decoupling corresponds to a semiclassical approximation mostly valid at high temperature, and neglects some of the coupling between the decoupled two-point function's correlation due to the imaginary time integration of the Kubo correlations.
A more formal decoupling keeping this quantum coupling is given by
\begin{equation}
\begin{split}
    \big(A B, C(t) D(t)\big) =& \kBT\int_0^\beta d\lambda \braket{A(i\hbar\lambda)B(i\hbar\lambda) C(t) D(t)} \\
    \approx& \kBT\int_0^\beta d\lambda \bigg[ \braket{A(i\hbar\lambda)C(t)}\braket{B(i\hbar\lambda) D(t)} \\
    &+ \braket{A(i\hbar\lambda)D(t)}\braket{B(i\hbar\lambda) C(t)} \\
    &+ \braket{A(i\hbar\lambda)B(i\hbar\lambda)}\braket{C(t) D(t)} \bigg]
\end{split}
\end{equation}
We can use the Fourier transform to express the first term of this equation in term of the standard correlation function
\begin{equation}
\begin{split}
    &\int_0^\beta d\lambda \braket{A(i\hbar\lambda) C(t)}\braket{B(i\hbar\lambda) D(t)} \\
    &= \int_{-\infty}^\infty d\omega_1d\omega_2 G_{AC}^<(\omega_1) G_{BD}^<(\omega_2) e^{-i(\omega_1 + \omega_2) t} \\
    &\times \int_0^\beta d\lambda e^{-\hbar (\omega_1 + \omega_2) \lambda} \\
    &= \int_{-\infty}^\infty d\omega_1d\omega_2 G_{AC}^<(\omega_1) G_{BD}^<(\omega_2) \frac{e^{\beta \hbar (\omega_1 + \omega_2)} - 1}{\hbar(\omega_1 + \omega_2)} \\
    &\times e^{-i(\omega_1 + \omega_2) t} \\
    &= \int_{-\infty}^{\infty} d\omega \frac{e^{\beta\hbar\omega} - 1}{\hbar\omega} e^{-i\omega t} \int_{-\infty}^{\infty} d\omega' G_{AC}^<(\omega') G_{BD}^<(\omega - \omega')
\end{split}
\end{equation}
Going further, we can also integrate this result from $t=0$ to infinity
\begin{equation}
\begin{split}
    &\int_0^\infty dt \braket{A(i\hbar\lambda) C(t)}\braket{B(i\hbar\lambda) D(t)} \\
    &= \int d\omega \frac{e^{\beta\hbar\omega} - 1}{\hbar\omega}  \int d\omega' G_{AC}^<(\omega') G_{BD}^<(\omega - \omega') \int_0^\infty dt e^{-i\omega t} \\
    &= \pi \int_{-\infty}^\infty d\omega' G_{AC}^<(\omega') G_{BD}^<(-\omega') \\
    &= \pi \int_{-\infty}^\infty d\omega' n(\omega') (n(\omega') + 1) \chi''_{AC}(\omega') \chi''_{BD}(\omega')
\end{split}
\end{equation}
where we used $G^<(\omega) = n(\omega) \chi''(\omega)$ and $G^<(-\omega) = G^>(\omega) = (n(\omega) + 1) \chi''(\omega)$ from appendix \ref{app:correlation functions}.

\section{Derivation of the Markovian approximation}
\label{app:markovian}

The Markovian approximation is founded on the assumption that the bath, represented by the memory kernel, follows a dynamic on a much slower time scale than the dynamical variable.
Effectively, this assumption translates into taking the infinite time limit of the convolution appearing in the generalized Langevin equation
\begin{equation}
    \int_0^t ds \Gamma_s(\vec{q}, s) \dot{G}_s(\vec{q}, t-s) \rightarrow \lim_{t\rightarrow\infty} \int_0^t ds \Gamma_s(\vec{q}, t-s) \dot{G}_s(\vec{q}, s)
\end{equation}
where we used the symmetry of convolutions $\int_0^t ds f(s) g(t-s) = \int_0^t ds f(t-s) g(s)$.
Furthermore, assuming that interactions with the bath are weak, the correlation function $\dot{G}_s(\vec{q}, t)$ in this equation can be approximated as acting as a delta function centered on $\Omega_s(\vec{q})$ in frequency space.
Using these approximations in the memory kernel convolution gives
\begin{equation}
\begin{split}
    &\int_0^\infty ds \Gamma_s(\vec{q}, t-s) \dot{G}_s(\vec{q}, s) \\
    &= \int_{-\infty}^{\infty} d\omega_1 d\omega_2 \Gamma_s(\vec{q}, \omega_1) G_s(\vec{q}, \omega_2) \omega_2 \int_0^\infty ds e^{i \omega_1 (t-s)} e^{i\omega_2 s} \\
    &= \int_{-\infty}^{\infty} d\omega \Gamma_s(\vec{q}, \omega) G_s(\vec{q}, \omega) \omega e^{i\omega t} \\
    &\approx \Gamma_s(\vec{q}, \Omega_s(\vec{q})) \int_{-\infty}^{\infty} d\omega G_s(\vec{q}, \omega) \omega e^{i\omega t} \\
    &= \Gamma_s^M(\vec{q}) \dot{G}_s(\vec{q}, t)
\end{split}
\end{equation}
where we defined the Markovian scattering rate $\Gamma_s^M(\vec{q}) = \Gamma_s(\vec{q}, \Omega_s(\vec{q}))$ found in the main text.

\section{Derivation of the scattering matrix}
\label{app:scattering matrix}

In this appendix, we derive the scattering matrix by computing the time integral of the four-point correlation function
\begin{equation}
    \big( A_{s_1}(\vec{q}_1)B_{s_1}(\vec{q}_1), A_{s_2}(\vec{q}_2, t)B_{s_2}(\vec{q}_2, t)\big)
\end{equation}
To facilitate the derivation, we introduce the composite operator
\begin{equation}
    C_{s}(\vec{q}, t) = A_s(\vec{q}, t) B_s(\vec{q}, t)
\end{equation}
whose correlation function is denoted by
\begin{equation}
    Y_{s_1,s_2}(\vec{q}_1,\vec{q}_2, t) = \big( C_{s_1}(\vec{q}_1), C_{s_2}(\vec{q}_2)\big)
\end{equation}
Our task is to compute the integral of this correlation function, which can be recognized as its zero-frequency Laplace transform.

From the definitions of the operator $A_s(\vec{q}, t)$ and $B_s(\vec{q}, t)$, and by projecting the Mori-Zwanzig equation of motion for $\dot{\vec{p}}$~\cite{Castellano2023} onto the phonon modes, we obtain the time derivative of these operators
\begin{align}
    \dot{A}_s(\vec{q}, t) =& -i\Omega_s(\vec{q})B_s(\vec{q}, t) \\
    \begin{split}
    \dot{B}_s(\vec{q}, t) =& i\Omega_s(\vec{q}) A_s(\vec{q}, t) \\
    &+ \frac{i}{\Omega_s(\vec{q})}\sum_{s'\vec{q}'}\int_0^t ds K_{s s'}(\vec{q}, \vec{q}', s) B_{s'}(\vec{q}', t-s) \\ &+ \frac{i}{\Omega_s(\vec{q})}\delta A_s(\vec{q}, t)
    \end{split}
\end{align}
The derivative of the composite operator $C_s(\vec{q}, t)$ can thus be expressed as
\begin{equation}
\begin{split}
    \dot{C}_s(\vec{q}, t) =& \dot{A}_s(\vec{q}, t)B_s(\vec{q}, t) + A_s(\vec{q}, t)\dot{B}_s(\vec{q}, t) \\
    =& -i\Omega_s(\vec{q}) B_s(\vec{q}, t) B_s(\vec{q}, t) + i\Omega_s(\vec{q}) A_s(\vec{q}, t)A_s(\vec{q}, t) \\
    &+ \frac{i A_s(\vec{q}, t) }{\Omega_s(\vec{q})}\sum_{s'\vec{q}'} \int_0^t ds K_{s s'}(\vec{q} \vec{q}', s) B_{s'}(\vec{q}', t-s) \\
    &- \frac{i}{\Omega_s(\vec{q})}\delta A_s(\vec{q}, t) A_s(\vec{q}, t)
\end{split}
\end{equation}
Multiplying by $C_{s}(\vec{q})$, taking the Kubo average and using the decoupling scheme, we obtain the following equation of motion for $Y_{s_1 s_2}(\vec{q}_1,\vec{q}_2, t)$
\begin{equation}
    \dot{Y}_{s_1 s_2}(\vec{q}_1 \vec{q}_2, t) = \sum_{s_3 \vec{q}_3}\int_0^t ds K_{s_2 s_3}(\vec{q}_2 \vec{q}_3, s) Y_{s_1 s_3}(\vec{q}_1 \vec{q}_3, t-s)
\end{equation}
or $\dot{\vec{Y}}(t) = \int_0^t ds \vec{K}(s) \vec{Y}(t-s)$ in matrix form.
From this equation of motion, we find that the Laplace transform of $\vec{Y}(t)$ can be written $\widetilde{\vec{Y}}(\omega) = (\omega - \widetilde{\vec{K}}(\omega))^{-1}\vec{Y}$, leading to the final result
\begin{equation}
    \widetilde{\vec{Y}}(0) = \widetilde{\vec{K}}(0)^{-1}\vec{Y}
\end{equation}
where $\widetilde{\vec{Y}}(\omega)$ and $\widetilde{\vec{K}}(\omega)$ are the Laplace transform of the four-point correlation function and the memory kernel, respectively.
Applying the decoupling rule for the $\vec{Y}$ matrix, it can be shown that it is a diagonal matrix with entries
\begin{equation}
    Y_{s s}(\vec{q},\vec{q}) = \Omega_s^2(\vec{q}) n_s(\vec{q}) (n_s(\vec{q}) + 1) = \kBT^2 c_s(\vec{q})
\end{equation}
Furthermore, we can recognize the scattering matrix of the main text as the Markovian limit of the memory kernel, including off diagonal terms
\begin{equation}
\begin{split}
    \Xi_{s_1 s_2}(\vec{q}_1, \vec{q}_2) =&  K_{s_1, s_2}(\vec{q}_1, \vec{q}_2, \Omega_{s_1}(\vec{q}_1)) \\
    =& \Gamma_{s_1}^M(\vec{q}_1) \delta_{\vec{q}_1 \vec{q}_2} \delta_{s_1 s_2} \\
    +&\frac{\Omega_{s_2}(\vec{q}_2)}{\Omega_{s_1}(\vec{q}_1)}\bigg[\sum_{\vec{q_3} s_3} [\Gamma_{s_1 s_2 s_3}^{(3)}(\vec{q}_1, \vec{q}_2, \vec{q}_3, \Omega_{s_1}(\vec{q}_1)) \\
    +& \Gamma_{s_1 s_3 s_2}^{(3)}(\vec{q}_1, \vec{q}_3, \vec{q}_2, \Omega_{s_1}(\vec{q}_1))] \\
    +& \sum_{\vec{q}_3 s_3 \vec{q}_4 s_4} [\Gamma_{s_1 s_2 s_3 s_4}^{(4)}(\vec{q}_1, \vec{q}_2, \vec{q}_3, \vec{q}_4, \Omega_{s_1}(\vec{q}_1)) \\
    +& \mathrm{perm}] \\
    +& \Gamma_{s_1 s_2}^{\mathrm{iso}}(\vec{q}_1, \vec{q}_2, \Omega_{s_1}(\vec{q}_1)) \bigg]
\end{split}
\end{equation}
where $\mathrm{perm}$ denotes all permutations of the second, third and fourth phonons in $\Gamma_{s_1 s_2 s_3 s_4}^{(4)}(\vec{q}_1, \vec{q}_2, \vec{q}_3, \vec{q}_4, \Omega_{s_1}(\vec{q}_1))$.

\section{Comparison with the phonon Boltzmann equation}
\label{app:pbe}

The central quantity in the phonon Boltzmann equation (PBE) is the phonon occupation  $n_s(\vec{q}, t)$~\cite{Cepellotti2016}.
To facilitate comparison with the results in the main text, we assume that the system is in a steady state with a homogeneous and constant temperature gradient.
Under these conditions, the spatial and temporal derivatives of the phonon occupation vanish, allowing us to eliminate the time dependence and work directly with $n_s(\vec{q})$.
Noting the equilibrium phonon occupation as $\bar{n}_s(\vec{q} = (e^{\beta \hbar\Omega_s(\vec{q})} - 1)^{-1}$, the Boltzmann equation is expressed as a balance between the temperature gradient and phonon scattering processes
\begin{equation}
    \frac{\partial \bar{n}_s(\vec{q})}{\partial T} \vec{v}_\vec{q}^s \nabla T = - \sum_{s'\vec{q}'}\Xi(\vec{q} s, \vec{q}', s) \delta n_{s'}(\vec{q}')
\end{equation}
where $\delta n_s(\vec{q}) = n_s(\vec{q}) - \bar{n}_s(\vec{q})$ represents the deviation of the phonon occupation from equilibrium, and  $\Xi(\vec{q}s, \vec{q}' s')$ is the phonon scattering matrix introduced in the main text.

Assuming a linearized form of the deviation,
\begin{equation}
    n_s(\vec{q}) = \bar{n}_s(\vec{q}) + \rho_s(\vec{q}) \nabla T
\end{equation}
the PBE reduces to
\begin{equation}
    \frac{\partial \bar{n}_s(\vec{q})}{\partial T} \vec{v}_\vec{q}^s = - \sum_{s'\vec{q}'}\Xi(\vec{q} s, \vec{q}', s) \boldsymbol{\rho}_{s'}(\vec{q}')
\end{equation}
Recognizing this as a matrix equation, we solve for $\rho_s(\vec{q})$ by inverting the scattering matrix:
\begin{equation}
    \rho_s(\vec{q}) = -\sum_{s'\vec{q}'} \frac{\partial n_{s}(\vec{q})}{\partial T}  \Xi^{-1}(s \vec{q}, s' \vec{q}') \vec{v}_{\vec{q}'}^{s'}
\end{equation}
where we used the fact that the scattering matrix is positive definite, hence symmetric.

To compute the thermal conductivity tensor, we introduce the heat current operator in the harmonic approximation,
\begin{equation}
    \boldsymbol{\mathcal{J}}^{\mathrm{harm}}(t) = -\frac{1}{V} \sum_{\vec{q},s} \hbar \Omega_s(\vec{q}) \vec{v}_\vec{q}^s \delta n_s(\vec{q}, t)
\end{equation}
Applying Fourier’s law, we express the thermal conductivity tensor as
\begin{equation}
\begin{split}
    \boldsymbol{\kappa}^{\mathrm{PBE}} =& \frac{\braket{\mathcal{J}^{\mathrm{harm}}}}{\nabla T} \\
    =& -\frac{1}{V} \sum_{\vec{q}, s} \hbar \Omega_s(\vec{q}) \vec{v}_\vec{q}^s \boldsymbol{\rho}_s(\vec{q})
\end{split}
\end{equation}
Substituting our solution for $\rho_s(\vec{q})$, we obtain
\begin{equation}
\begin{split}
    \boldsymbol{\kappa}^\mathrm{PBE} =& -\frac{1}{V} \sum_{s \vec{q}}\sum_{s' \vec{q}'} \hbar\Omega_s(\vec{q}) \frac{\partial n_s(\vec{q})}{\partial T} \vec{v}_\vec{q}^s \otimes \vec{v}_{\vec{q}'}^{s'} \Xi^{-1}(s\vec{q}, s'\vec{q}') \\
    =& \frac{1}{V} \sum_{s \vec{q}}\sum_{s' \vec{q}'} c_s(\vec{q}) \vec{v}_\vec{q}^s \otimes \vec{v}_{\vec{q}'}^{s'} \Xi^{-1}(s\vec{q}, s'\vec{q}')
\end{split}
\end{equation}
which coincide with eq.(\ref{eq:kappa diag bte}) of the main text.

\section{Treatment of degeneracies}
\label{app:degen}

In reciprocal space, the generalized group velocities can be seen through Hellman-Feynman theorem as the derivative of the dynamical matrix projected on phonons $s$ and $s'$ of the same $\vec{q}$-point.
We can write this formally as
\begin{equation}
\begin{split}
    \vec{v}_\vec{q}^{ss'} = \frac{1}{2\sqrt{2\Omega_s(\vec{q})\Omega_{s'}(\vec{q})}} \bra{s,\vec{q}} \nabla_\vec{q} \vec{\Phi}(\vec{q}) \ket{s',\vec{q}}
\end{split}
\end{equation}
where we introduced the notation
\begin{equation}
    \bra{s,\vec{q}} O \ket{s',\vec{q}} = \sum_{i\alpha,j\beta} \varepsilon_s^{i\alpha *}(\vec{q}) \varepsilon_{s'}^{j\beta}(\vec{q}) O_{i\alpha}^{j\beta}
\end{equation}
where $\vec{O}$ is a $3 N_\mathrm{at}\times 3 N_\mathrm{at}$ matrix.
For degenerate modes, a direct calculation with this formula is ill defined and we use instead degenerate perturbation theory.
We start by computing the $N_\mathrm{degen}\times N_\mathrm{degen}$ matrix $\vec{h}$ composed of the derivative of the dynamical matrix only for modes $s_i$ and $s_j$ in the degenerate subspace
\begin{equation}
\begin{split}
    h_{ij}^\alpha = \bra{s_i,\vec{q}} \frac{\partial}{\partial q^\alpha} \boldsymbol{\Phi}(\vec{q}) \ket{s_j, \vec{q}}
\end{split}
\end{equation}
Then, the diagonal component of the generalized group velocities for each of the degenerate modes are computed as the average of the eigenvalues $\lambda_i$ of $\vec{h}$
\begin{equation}
    v_{\vec{q}}^{s,\alpha} = \frac{1}{2\Omega_s(\vec{q})} \frac{1}{N_\mathrm{degen}} \sum_i \lambda_i
\end{equation}
while the off-diagonal components vanish:
\begin{equation}
    v_{\vec{q}}^{s_i s_j,\alpha} = 0
\end{equation}

\section{Accuracy of the Machine-Learning Interatomic Potentials}
\label{app:param}

In this appendix, we give the computational details for the fitting of the machine-learning interatomic potential used in the applications of section \ref{sec:applications}.
In all cases, the MLIP were constructed by fitting on \textit{ab initio} data computed with DFT using the Abinit package~\cite{Gonze2020,Romero2020}, with the dataset created following the scheme presented in section \ref{sec:framework}.
The functional, \textbf{k}-point grid and kinetic energy cutoff used for each system are detailed in table \ref{tab:dft params}, with the parameters selected to give a total energy accuracy better than 1meV/atom.
%For PbTe, we included spin-orbit interactions in our DFT calculation since relativistic effects are known to be important for a good description of vibrational properties.
For Li\textsubscript{3}ClO and AgI, Born effective charges and dielectric constant were computed at the ground state volumes using DFPT as implemented in Abinit.
The non-analytical long-range corrections were then applied using the method described in the supplementary materials of \cite{Zhou2018}.

On fig.\ref{fig:correlations}, we show the correlations between DFT and MLIP energy, forces and stress for each system while table \ref{tab:dft vol} compares ground state lattice parameters and volume.
For all materials studied here, the agreement is excellent, both for the error measure given by the root-mean squared error and the mean absolute error and the comparison of the lattice parameters.

\begin{figure*}
    \centering
    \includegraphics[width=\linewidth]{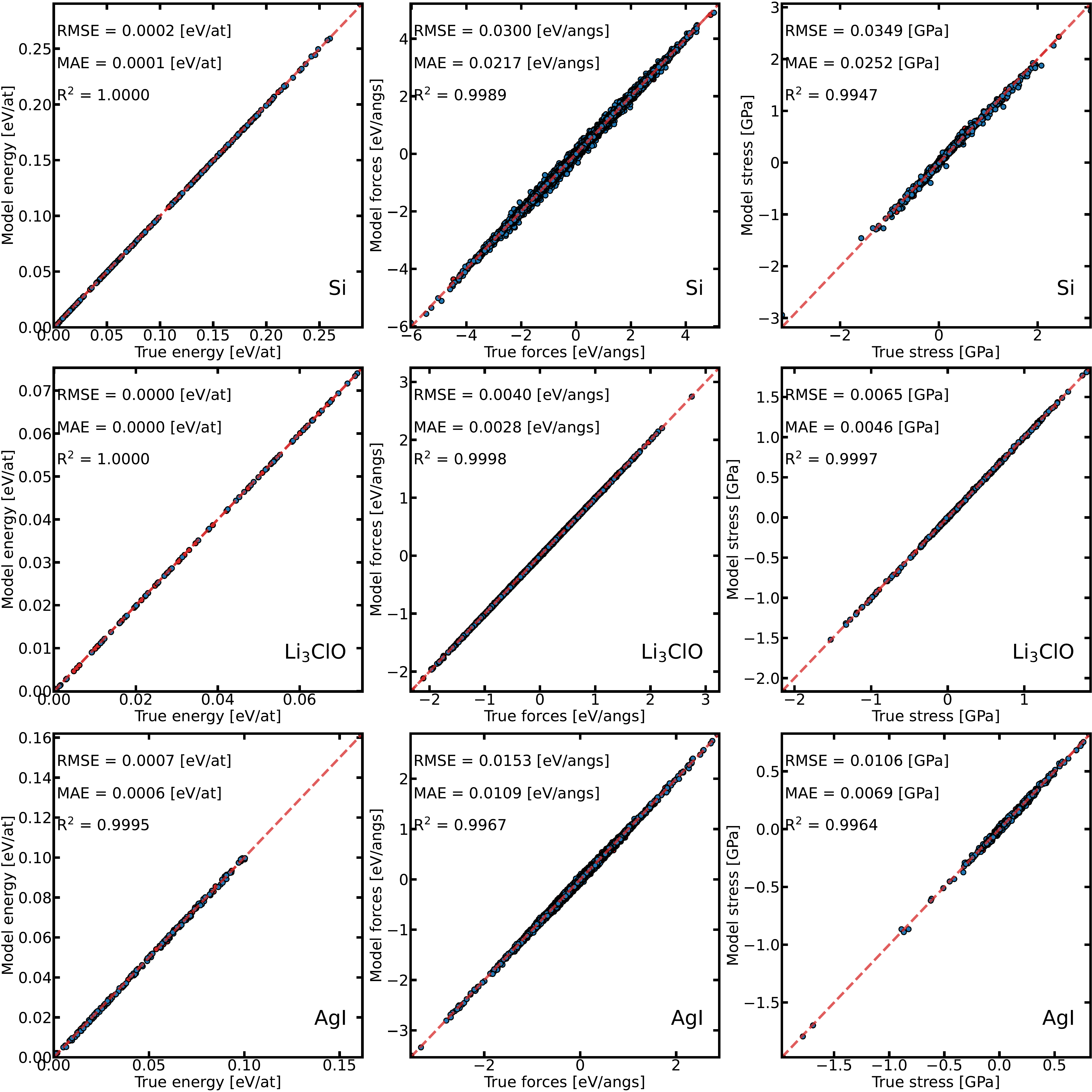}
    \caption{Correlation between DFT and MLIP energy, forces and stress for all the MLIP used in this work.}
    \label{fig:correlations}
\end{figure*}

\begin{table}[h]
    \centering
    \begin{ruledtabular}
    \begin{tabular}{c|ccc}
        system & XC functional & \textbf{k}-point grid & cutoff [Ha] \\
        \hline
         Si & PBEsol\footnotemark[1]~\cite{Perdew2008} & $12\times12\times12$ & 25 \\
%        Graphene & PBE\footnotemark[1] & $18\times18$ & 45 \\
         Li\textsubscript{3}ClO & PBE\footnotemark[2] & $8\times8\times8$ & 32\\
%        PbTe & PBE\footnotemark[2] & $12\times12\times12$ & 22 \\
         AgI & PBE\footnotemark[2] & $8\times8\times8$ & 32 \\
    \end{tabular}
    \end{ruledtabular}
    \footnotetext[1]{Norm-conserving, from pseudodojo\cite{vanSetten2018}}
    \footnotetext[2]{PAW, from GPAW pseudopotential dataset~\cite{Mortensen2024}}
    \caption{Parameters for the DFT calculations for each of the systems used in the applications.}
    \label{tab:dft params}
\end{table}

\begin{table}[h]
    \centering
    \begin{ruledtabular}
    \begin{tabular}{c|cccc}
        system & \multicolumn{2}{c}{DFT} & \multicolumn{2}{c}{MLIP} \\
        & a [\AA] & volume [\AA\textsuperscript{3}] & a [\AA] & volume [\AA\textsuperscript{3}] \\
        \hline
        Si  & 5.431 & 20.026 & 5.431 & 20.026\\
%       Graphene & 2.466 & & 2.466 \\
        Li\textsubscript{3}ClO & 3.916 & 12.010 & 3.916 & 12.010 \\
        AgI & 6.654 & 36.834 & 6.654 & 36.828 \\
    \end{tabular}
    \end{ruledtabular}
    \caption{Comparison between the ground state lattice constant computed with DFT and the MLIP.}
    \label{tab:dft vol}
\end{table}

\section{Approach to equilibrium molecular dynamics}
\label{app:aemd}

The approach-to-equilibrium molecular dynamics (AEMD)~\cite{Sheng2022,Lampin2013,Martin2022,Lambrecht2024,Melis2014} method directly applies Fourier's law, utilizing the time evolution of temperature governed by the heat equation to estimate the thermal conductivity.
In this approach, the simulation box is initially divided into two regions : a hot region and cold region, thermostated at temperatures $T + \Delta T / 2$ and $T - \Delta T / 2$, respectively.
After the regions reaches constrained equilibrium, the thermostats are removed, allowing the system to relax towards a global equilibrium.
During this relaxation, the time evolution of the temperature difference between the two regions is described by the heat equation
\begin{equation}
\label{eq:fourier aemd}
    \Delta T(t) = \sum_{m=0}^\infty \frac{8 \Delta T(0)}{(2 m + 1)^2 \pi^2} e^{-(2 m + 1)^2 t / \tau}
\end{equation}
where $\tau$ is the decay time, which is related to the system's thermal conductivity through the equation
\begin{equation}
    \kappa(L) = \frac{L C_v}{4\pi^2 S}\frac{1}{\tau}
\end{equation}
Here, $L$ is the length of the system along the temperature gradient, $C_v$ is the heat capacity, and $S$ is the cross-sectional area perpendicular to the heat flow.

As a real-space method, AEMD is subject to finite-size effects.
Recently, Sheng \textit{et al}~\cite{Sheng2022} used lattice dynamics analysis to derive a model that fits and extrapolates the size dependence of thermal conductivity from AEMD.
Their model is expressed as
\begin{equation}
\label{eq:aemd extrapolate}
    \hat{\kappa}(L) = \frac{\kappa_1}{1 + \bigg(\frac{2\pi \Lambda_1}{L}\bigg)^2} + \frac{\kappa_2}{1 + \bigg(\frac{2\pi \Lambda_2}{L}\bigg)^2}
\end{equation}
where $\kappa_1$, $\kappa_2$, $\Lambda_1$ and $\Lambda_2$ are fitting parameters.

In our simulations, we applied a temperature difference of $\Delta T=200$~K.
To fit the time evolution of the temperature difference, using 10 terms of equation (\ref{eq:fourier aemd}) was sufficient to accurately converge the decay time $\tau$.
For each system and temperature, thermal conductivity was computed for various systems length $L$, and the results were extrapolated to the bulk limit using equation (\ref{eq:aemd extrapolate}).
The length of the system along the temperature gradient were $273$, $546$, $1092$, $2731$, $3278$, $5463$ and $8156$~\r{A} for Si, $40$, $199$, $398$ and $795$~\r{A} for Li\textsubscript{3}ClO and $334$, $1337$, $2674$ and $4679$~\r{A} for AgI.
Our simulations were run for several hundred picoseconds, depending on the length of the system, with a timestep of 1 fs and data gathered every 50 fs.

\bibliography{biblio}% Produces the bibliography via BibTeX.

\end{document}